\documentclass{article}
\usepackage{epsfig}

\date{\today}

\newcommand{\ka}{\kappa}
\newcommand{\al}{\alpha}

\newcommand{\si}{\sigma}

\newcommand{\ee}{\end{equation}}
\newcommand{\eea}{\end{eqnarray}}
\newcommand{\be}{\begin{equation}}
\newcommand{\bea}{\begin{eqnarray}}

\newcommand{\re}[1]{(\ref{#1})}

\begin{document}

\begin{center}
{\Large\bf No hair conjecture, nonabelian hierarchies \\
and anti-de Sitter spacetime}
\vspace{0.5cm}
\\
{\bf Eugen Radu$^{1}$ }
{\bf and D. H. Tchrakian $^{1,2}$ }

\vspace*{0.2cm}
{\it $^1$ Department of
Mathematical Physics, National University of Ireland Maynooth, Maynooth,
Ireland}

{\it $^{2}$School of Theoretical Physics -- DIAS, 10 Burlington
Road, Dublin 4, Ireland}
\vspace{0.5cm}
\end{center}
\begin{abstract}
We consider globally regular and black holes solutions for the
Einstein-Yang-Mills system with negative cosmological constant in $d-$spacetime
dimensions.
We find that the ADM mass of the spherically symmetric solutions generically
diverges for $d>4$.
Solutions with finite mass are found by considering
corrections to the YM Lagrangean consisting in higher therm of the  
Yang--Mills hierarchy.
 Such systems can occur in the
low energy effective action of string theory.
A discussion of the main properties
of the solutions and the differences with respect to the four dimensional case
is presented.
The mass of these configurations is computed by using a counterterm method.
\end{abstract}

%%%%%%%%%%%%%%%%%%%%%%%%%%%%%%%%%%%%%%%%%%%%%%%%%%%%%%%%%%%%%%%%
\section{Introduction}
%%%%%%%%%%%%%%%%%%%%%%%%%%%%%%%%%%%%%%%%%%%%%%%%%%%%%%%%%%%%%%%%
If we allow for a negative cosmological constant, 
the solution of the matter free Einstein equations possessing
the maximal number of symmetries is the anti-de Sitter (AdS) spacetime.
Recently a considerable amount of interest has been focused on solutions of
Einstein equations with this type of asymptotics. This interest is mainly
motivated by the proposed correspondence between physical effects 
associated with gravitating fields propagating in AdS spacetime and those of a
conformal field theory on the boundary of AdS spacetime \cite{Maldacena:1997re,
Witten:1998qj}.

In view of these developments, an examination of the classical solutions of
gravitating  fields in asymptotically AdS (AAdS) spacetimes seems appropriate.
Recently, some authors have discussed the properties  of gravitating SU(2)
nonabelian fields with a negative cosmological constant $\Lambda$
\cite{Winstanley:1998sn,Bjoraker:2000qd}. Considering the case of four
spacetime dimensions, they obtained some surprising results, which are
strikingly different from  the results found in the asymptotically flat case.
For example there are solutions for continuous intervals of the parameter
space,  rather then discrete points. The asymptotic values of the gauge
potentials are arbitrary and there exist solutions supporting magnetic and
electric fluxes without a Higgs field. Some of these solutions are
stable against spherically symmetric linear perturbations. The literature on
AAdS solutions with nonabelian fields is growing steadily,
including stability analyses \cite{Sarbach:2001mc,Breitenlohner:2003qj}, 
the study of configurations with a NUT charge \cite{Radu:2002hf},
topological black holes with nonabelian hair \cite{VanderBij:2001ia} as well as
axially symmetric generalisations \cite{Radu:2002rv,Radu:2004gu}. The existence
of these solutions invalidates the AdS$_4$ version of the no hair conjecture, 
which states that the black holes are completely characterised
by their mass, charge and angular momentum.

However, all these studies approach the case of a four dimensional AAdS
spacetime, and relatively little is known about higher dimensional AAdS
solutions with non Abelian matter fields. Practically all that is known for
$d>4$ is the five dimensional nonabelian $SU(2)$ solutions discussed in
\cite{Okuyama:2002mh}. At the same time gauged supergravity theories playing
an important role in AdS/CFT, generically contain non Abelian matter fields in
the bulk, although in the literature only Abelian truncations are considered,
to date. Thus, the examination of higher dimensional gravitating
non Abelian solutions with $\Lambda<0$ is a pertinent task.

Higher dimensional asymptotically flat solutions of the Einstein-Yang-Mills
(EYM) equations have recently been the subject of several studies.
As found in \cite{Volkov:2001tb}, in five spacetime dimensions,
the particle spectrum
obtained by uplifting the $d=4$ flat space YM instantons become 
completely destroyed by gravity, as a result of their scaling behaviour.
However, by adding higher order~\footnote{The only higher order curvature terms
considered in this paper are those constructed from a $2p$-form field such that
the Lagrangean contains {\it velocity square} fields only.}
terms in the Yang-Mills (YM) hierarchy this
obstacle due to scaling is removed and solutions in higher dimensions
can be found. Such regular, static and spherically symmetric solutions in
spacetime dimensions $d\le 8$ were
presented in \cite{Brihaye:2002hr}, and for $d=5$, both
globally regular and black hole solutions 
of the EYM system were found in \cite{Brihaye:2002jg}.
The properties of these solutions are rather different from the familiar
Bartnik-McKinnon (BK) solutions \cite{Bartnik:1988am} to EYM in $d=4$, and are
somewhat more akin to the gravitating monopole solutions to EYM-Higgs (EYMH)
\cite{Breitenlohner:1992}. This is because like in the latter case
\cite{Breitenlohner:1992}, where the VEV of the Higgs field features as an
additional dimensional constant, here \cite{Brihaye:2002hr} also additional
dimensional constants enter with each higher order YM curvature
term~\footnote{Like the gravitating monopoles, these have gravity decoupling
limits except in $d=5$ (and in $d=4p+1$ modulo $4$), and in all {\it odd}
spacetime dimensions, the flat space solutions are stabilised by a
Pontryagin charge analogous to the magnetic charge of the monopole, provided
that the representations of the gauge group are chosen suitably. In all
{\it even} $d$ however, they are like the BK solution in that they are not
stabilised by a topological charge and are likewise sphalerons
\cite{Brihaye:2004ff}}. They
are however quite distinctive in their critical behaviour. The typical
critical features discovered in \cite{Brihaye:2002hr,Brihaye:2002jg} have
recently been analysed and explained in \cite{Breitenlohner:2005}.

These results can be systematically extended to all $d\ge 5$ and one finds that
no finite mass solutions can exist in EYM theory, unless one modifies the 
non Abelian action by adding higher order curvature terms~\footnote{In
principle higher order terms with the desired scaling can be chosen
to consist both of the YM and the Riemann curvatures, but in practice
we restrict to the YM hierarchy. The reason will be explained in section
{\bf 2.5}. Besides, it was found in \cite{Brihaye:2002hr} that the inclusion of
Gauss-Bonnet terms does not result in any new qualitative features to the
solutions.} in the YM $2p$-form curvature $F(2p)$, the $p=1$ case being the
usual $2$-form YM curvature.
Without these higher order YM terms, only vortex-type finite energy solutions
\cite{Volkov:2001tb} exist, describing effective systems
in $3$ spacelike dimensions, and with a number of codimensions. 

It is the purpose of this paper to examine the corresponding situation in
higher dimensions, 
in the presence of a negative cosmological constant. It is interesting to
inquire whether the introduction of a negative cosmological constant to
these higher dimensional EYM models will lead to some new effects as it does in
the $d=4$ case, due to the different asymptotic structure of the spacetime. 
In the first place this would lead to our understanding of how the behaviour
of EYM theory depends on the dimensionality of the spacetime. But such higher
dimensional AAdS solutions might be relevant to superstring
theory, namely in the context of solutions to various supergravities
containing nonabelian matter fields. Here
however we restrict our considerations to the simplest case of systems
consisting only of gravitational and YM fields, namely to
higher dimensional EYM model. In particular, we restrict to
Einstein-Hilbert gravity and the first two 
members of the YM hierarchy, and hence to $d\le 8$, augmented with negative
cosmological constant in $d$ spacetime dimensions. As in
\cite{Winstanley:1998sn,Bjoraker:2000qd}, for the $d=4$ case, we seek static
spherically symmetric solutions in the $d-1$ spacelike dimensions. We
find both globally regular and black hole solutions with finite ADM mass.
Unlike in the $d=4$ case however, we find that for $d>4$ and a negative
cosmological constant, the properties of the AAdS solutions do not differ
qualitatively from the asymptotically flat case.

Our strategy is to first consider the usual YM model, namely the $p=1$ member
of the YM hierarchy, or the square of the $2$-form curvature $F(2)$.
We present an argument for the absence of
 solutions with reasonable asymptotics
for any spacetime dimension $d\geq 5$. Although the EYM
equations in this case present solutions approaching asymptotically the AdS
background, the mass generically diverges.
This can be seen as a simple version of the no hair theorem, holding for the
EYM system in $d>4$ dimensions. In other words, Schwarzschild-AdS black hole
is the only static, spherically symmetric solution of the EYM system with
finite mass. This is presented in Section 2, where in addition we have
considered the special case $d=3$, extending Deser's analysis
\cite{Deser:1984fk} for $\Lambda<0$.

In Section 3 we introduce the higher order YM hierarchy models featuring the
terms $F(2p)$, $p\ge 2$. We
derive the classical equations subject to our spherically symmetric ansatz, 
and present a detailed  numerical study of of both regular and black hole
solutions. As is the case with the usual EYM system, where the existence of
finite mass regular solutions leads to the violation of the no hair conjecture,
here too, these solutions are explicit counterexamples to this
conjecture in AdS$_d$ spacetime.

One may ask about the possible relevance of
these higher dimensional configurations within the AdS/CFT correspondence.
In Section 4 we compute the boundary stress tensor and the 
mass and action of the solutions
in a number of spacetime dimensions up to eight. In five and seven
dimensions, the counterterm prescription of \cite{Balasubramanian:1999re}
gives an additional vacuum  (Casimir) energy, which agrees with that found in
the context of AdS/CFT correspondence. A counterterm based proposal to remove
the divergences of a $F(2)$ theory, such that the mass and action be
finite, is also presented in Appendix B.
We give our conclusions and remarks in the final section.

Everewhere in this paper we employ the notations and conventions of 
\cite{Brihaye:2002hr}.
%%%%%%%%%%%%%%%%%%%%%%%%%%%%%%%%%%%%%%%%%%%%%%%%%%%%%%%%%%%%%%%%%%%%%%%%%%%%%%
\section{The $F(2)$ model}
%%%%%%%%%%%%%%%%%%%%%%%%%%%%%%%%%%%%%%%%%%%%%%%%%%%%%%%%%%%%%%%%%%%%%%%%%%%%%%
%%%%%%            general action
%%%%%%%%%%%%%%%%%%%%%%%%%%%%%%%%%%%%%%%%%%%%%%%%%%%%%%%%%%%%%%%%%%%%%%%%%%%%%%
\subsection{The action principle}
We start with the following action principle in $d-$spacetime dimensions
\begin{eqnarray}
\label{action}
I=\int_{\mathcal{M}} d^d x \sqrt{-g}
\left(
 \frac{1}{16 \pi G}(R-2 \Lambda)
 +{\cal L}_m
\right)
-\frac{1}{8 \pi G}\int_{\partial\mathcal{M}} d^3 x\sqrt{-h}K,
\end{eqnarray}
where $R$ is the Ricci scalar associated with the
spacetime metric $g_{\mu\nu}$,  $\Lambda=-(d-1)(d-2)/(2 \ell^2)$ is the
cosmological constant and $G$ is the gravitational constant 
(following \cite{Brihaye:2002jg}, we define also $\kappa=1/(8\pi G)$).
 
The matter term in the above relation
\begin{eqnarray}
\label{F2}
{\cal L}_m=-\frac{1}{4} \tau_1 ~{\rm tr}~F_{\mu \nu} F^{\mu \nu}
\end{eqnarray}
is the usual $F^2$ nonabelian action density,
$F_{\mu\nu}=\partial_\mu A_\nu-\partial_\nu A_\mu-i[A_\mu, A_\nu]$ being the gauge field 
strength tensor.  
Here $\tau_1$ is the coupling constant of the model (in the usual notation
$\tau_1=1/g^2$).

The last term in  (\ref{action}) is the Hawking-Gibbons surface term
\cite{Gibbons:1976ue},  where $K$ is the trace 
of the extrinsic curvature for the boundary $\partial\mathcal{M}$ and $h$ is
the induced  metric of the boundary. Of course this term does not affect the
equations of motion but it is relevant for the discussion of the mass and the
action of the solutions, in Section 4.

The field equations are obtained by varying the action (\ref{action}) with
respect to the field variables $g_{\mu \nu},A_{\mu}$ 
\begin{eqnarray}
\label{Einstein-eqs}
R_{\mu \nu}-\frac{1}{2}g_{\mu \nu}R +\Lambda g_{\mu \nu}&=&
8\pi G  T_{\mu \nu},
\\
\nonumber
\nabla_{\mu}F^{\mu\nu}-i[A_{\mu},F^{\mu\nu}]&=&0,
\end{eqnarray}
where the energy momentum tensor is defined by
\begin{eqnarray}
\label{Tij}
T_{\mu\nu} =
    {\rm tr}~F_{\mu\alpha} F_{\nu\beta} g^{\alpha\beta}
   -\frac{1}{4} g_{\mu\nu} {\rm tr}~ F_{\alpha\beta} F^{\alpha\beta}.
\end{eqnarray}
%%%%%%%%%%%%%%%%%  line element %%%%%%%%%%%%%%%%%
\subsection{The general Ansatz}
%%%%%%%%%%%%%%%%%%%%%%%%%%%%%%%%%%
For the case of a $d$-dimensional spacetime, we restrict to static fields that
are spherically symmetric in the $d-1$ spacelike dimensions, with a
metric ansatz in terms of Schwarzschild coordinates
\begin{eqnarray}
\label{metric}
ds^{2}=\frac{dr^2}{N(r)}+r^{2}d \Omega_{d-2}^2-\sigma^2(r)N(r)dt^2,
\end{eqnarray}  
with  $d\Omega_{d-2}$ the $d-2$ dimensional 
angular volume element and
\begin{eqnarray}
\label{N}
N=1-\frac{2m(r)}{\kappa r^{d-3}}+\frac{r^2}{\ell^2},
\end{eqnarray}
the function $m(r)$ being related to the local mass-energy density up to some
$d-$dependent factor.

As discussed in \cite{Brihaye:2002hr}, the choice of gauge group 
compatible with the symmetries of the line element (\ref{metric}) 
is somewhat flexible. 
In \cite{Brihaye:2002hr} the gauge
group chosen was $SO(d)$, in $d$ dimensions. But the gauge field of
the static solutions in question took their values in $SO(d-1)$. Thus in
effect, it is possible to choose $SO(d)$ in the first place. Now for {\it
even} $d$, it is convenient to choose $SO(d)$ since we can then avail
of the chiral representations of the latter, although this is by no means
obligatory. Adopting this criterion, namely to employ chiral
representations, also for {\it odd} $d$, it is convenient to choose the
gauge group to be $SO(d-1)$. We shall therefore denote our representation
matrices by $SO_{\pm}(\bar d)$, where $\bar d=d$ and $\bar d=d-1$ for
{\it even} and {\it odd} $d$ respectively.

In this unified notation (for odd and even $d$), the spherically symmetric
Ansatz for the $SO_{\pm}(\bar d)$-valued gauge fields then reads
\cite{Brihaye:2002hr}
\begin{equation}
\label{YMsph}
A_0=0\ ,\quad
A_i=\left(\frac{1-w(r)}{r}\right)\Sigma_{ij}^{(\pm)}\hat x_j\ , \quad
\Sigma_{ij}^{(\pm)}=
-\frac{1}{4}\left(\frac{1\pm\Gamma_{ \bar d+1}}{2}\right)
[\Gamma_i ,\Gamma_j]\ .
\end{equation}
The $\Gamma$'s denote the $\bar d$-dimensional gamma matrices and
$1,~j=1,2,...,d-1$ for both cases.

Inserting this ansatz into the
action (\ref{action}), the EYM field equations reduce to
\begin{eqnarray}
\label{eqsF2-1}
0&=&\left(r^{d-4}\si Nw'\right)'
-(d-3)r^{d-6}\si(w^2-1)w,
\\
\label{eqsF2-2}
m' &=&\frac{\tau_1}{2}r^{d-4}\left( Nw'^2+(d-3)\frac{(w^2-1)^2}{2r^2}\right),
\\
\label{eqsF2-3}
\frac{  \sigma'}{\sigma}  &=&\frac{ \tau_1 }{\kappa } \frac{w'^2}{r},
\end{eqnarray}
which can also be derived from the reduced action
\begin{eqnarray}
\label{eff-action-F2}
S=\int ~dr~
\sigma \left(m'-\frac{\tau_1}{2}r^{d-4}\left( Nw'^2+(d-3)\frac{w^2-1)^2}{2r^2}
\right)\right).
\end{eqnarray}
For a $F^2$ theory,
the constants $\kappa,~\tau_1$ can always be absorbed by rescaling $r \to cr$, 
$\Lambda \to \Lambda/c^2$ 
and $ m \to m \kappa c^{d-3}$, with $c=\sqrt{\tau_1/\kappa}$.

The above differential equations have two analytic solutions. One of them with
\begin{eqnarray}
w(r)=\pm 1,~~m(r)=M,~~\sigma(r)=1
\end{eqnarray}
corresponds to Schwarzschild-AdS spacetime.
For $w(r)=0$ we find a non Abelian generalisation of the magnetic-
Reissner-Nordstr{\o}m solution with $\sigma(r)=1$ and
\begin{eqnarray}
\label{RN-sol}
m(r)=M_0+\frac{\tau_1}{2}\log r ~~{\rm if} ~~d=5, ~~{\rm and} ~~
~~m(r)=M_0+\frac{\tau_1(d-3)}{4(d-5)}r^{d-5}
~~{\rm for} ~~d\neq 5, 
\end{eqnarray}
$M_0$ being an arbitrary constant.
We can see that, although these solutions are asymptotically AdS, 
the mass defined in the usual way diverges.
%%%%%%%%%%%%%%%%%%%%%%%%%%%%%%%%%%%%%%%%%%%%%%%%%%%%%%%%%%%%%%%%%%%%%%%%%%%%%%
\subsection{$d=3$ }
%%%%%%%%%%%%%%%%%%%%%%%%%%%%%%%%%%%%%%%%%%%%%%%%%%%%%%%%%%%%%%%%%%%%%%%%%%%%%%
The (2+1) dimensional case is rather special.
Three dimensional gravity has provided 
many important clues about higher dimensional physics.
This theory with $\Lambda<0$ has non-trivial solutions, such as the BTZ
black-hole spacetime \cite{Banados:wn}, which provide an important testing
ground for quantum gravity and the AdS/CFT correspondence.
Many other types of $3d$ regular and black hole solutions with a negative
cosmological constant have also been found by coupling matter fields to gravity
in different ways.
 
However, as proven in \cite{Deser:1984fk}, 
there are no $d=3$ asymptotically flat static solutions of the 
EYM equations.
The arguments in \cite{Deser:1984fk} can easily be generalised
for the AAdS case. We notice that for $d=3$, the YM equation (\ref{eqsF2-1})
implies the existence of a first integral $w'=\alpha r/(\sigma N)$ with 
$\alpha$ an arbitrary real constant. Therefore, assuming
AdS$_3$ asymptotics, $w'$ decays asymptotically as $1/r$ which from 
(\ref{eqsF2-2}) implies a divergent value of $m(r)$ as $r \to \infty$.
However, similar to the $\Lambda=0$ case \cite{maison}, this argument does not
exclude the existence of nontrivial solutions of the field equations.

Here we should remark that since 
for $d=3$ we are dealing with $SO(d-1) = SO(2)$, the gauge group is
Abelian and we recover Einstein-Maxwell theory with a negative
cosmological constant, whose solutions are known in the literature.
The corresponding solution with a vanishing electric
field   was found by Hirschmann and Welch  \cite{Hirschmann:1995he} 
and has a line element
\footnote{One can also solve directly the field equations (\ref{eqsF2-1})-(\ref{eqsF2-3}),
but the solution takes a much more complicated form for the metric ansatz (\ref{metric}).}
\begin{eqnarray}
\label{H1}
ds^2=\frac{r^2}{ r^2+c^2 \log |r^2/\ell^2-M|}
\frac{dr^2}{(r^2/\ell^2-M)}+(r^2+c^2\log |r^2/\ell^2-M|)d\varphi^2
-(r^2/\ell^2-M)dt^2,
\end{eqnarray}
the magnetic potential being
\begin{eqnarray}
\label{H2}
w(r)=w_0+\frac{c}{\sqrt{2}}\log |r^2/\ell^2-M|,
\end{eqnarray}
with $w_0,~c$ and $M$ arbitrary real constants, the BTZ metric being recovered for $c=0$  
(see also  \cite{Cataldo:1996ue, Kiem:1996ks} for more details on this solution).

One can see that although the quasilocal mass defined in the
usual way diverges as $r \to \infty$, the metric still approaches 
the AdS$_3$ background. However, a similar problem appears for other $d=3$ AdS solution
$e.g.$  for the electrically charged BTZ black hole \cite{Banados:wn}, 
or for a self-interacting scalar
field minimally coupled to gravity \cite{Henneaux:2002wm},
in which cases it was possible to find a suitable mass definition.
We expect the formalism developed in those cases to work also for
the Hirschmann-Welch solution (\ref{H1})-(\ref{H2}), but this lies
outside the scope of the present work.
 
%
%%%%%%%%%%%%%%%%%%%%%%%%%%%%%%%%%%%%%%%%%%%%%%%%%%%%%%%%%%%%%%%%%%%%%%%%%%%%%%
\subsection{$d=4$}
%%%%%%%%%%%%%%%%%%%%%%%%%%%%%%%%%%%%%%%%%%%%%%%%%%%%%%%%%%%%%%%%%%%%%%%%%%%%%%
Four dimensional black hole solutions of the equations
(\ref{eqsF2-1})-(\ref{eqsF2-3}) have been found in \cite{Winstanley:1998sn},
the  globally regular counterparts being discussed in  \cite{Bjoraker:2000qd}.
Differing from the asymptotically flat case, for $\Lambda<0$
there is a continuum of solutions in terms of the adjustable shooting parameter
that specifies the initial conditions at the origin or at the event horizon.
As a new feature, the asymptotic value of the gauge function $w_0$ is
arbitrary. The spectrum has a finite number of continuous branches, depending
on the value of $\Lambda$. When the parameter $\Lambda$ approaches zero, 
an already-existing branch of  solutions collapses to a single point in the
moduli space. At the same time new branches of solutions emerge.
A fractal structure in the moduli space has been noticed
\cite{Hosotani:2001iz}. 
There are also nontrivial solutions stable against spherically symmetric 
linear perturbations, corresponding  to stable configurations.
The solutions are classified  by non-Abelian magnetic charge and the ADM mass.

Note also that the $d=4$ EYM solutions with a negative cosmological constant
$\Lambda=-3/\tau_1$ have some relevance in AdS/CFT context.
As proven in \cite{Pope:1985bu}, for this value of the cosmological constant,
an arbitrary solution $(g_{\mu \nu},A_{\mu}^{(a)})$ of the four dimensional
EYM equations gives a solution of the equations of motion of the $d=11$
supergravity. Based on this observation, an exact BPS-type EYM solution
has been constructed in \cite{Radu:2004gu}. However, similar to some
supersymmetric solutions in Einstein-Maxwell theory with $\Lambda<0$, 
this $\Lambda=-3/\tau_1$ configuration presents a naked central singularity.

%%%%%%%%%%%%%%%%%%%%%%%%%%%%%%%%%%%%%%%%%%%%%%%%%%%%%%%%%%%%%%%%%%%%%%%%%%%%%%
\subsection{$d\geq 5$ }
%%%%%%%%%%%%%%%%%%%%%%%%%%%%%%%%%%%%%%%%%%%%%%%%%%%%%%%%%%%%%%%%%%%%%%%%%%%%%%
As discovered by Coleman \cite{coleman} and Deser \cite{Deser:1976wq},  
there are no flat space static solutions of the YM equations, except for $d=5$.
However, the inclusion of gravity may change this picture, as seen from the 
famous $d=4$ asymptotically flat Bartnik-McKinnon solutions
\cite{Bartnik:1988am}. In this case, the repulsive 
YM force is compensated by the attractive character of the gravity,
and as a result we find both regular and black hole unstable configurations 
(see \cite{Volkov:1998cc} for a fairly recent survey).
As found in \cite{Volkov:2001tb} the $d=5$ particle-like 
solutions are destroyed by gravity, their mass diverging logarithmically, while
$w(r)$ presents an infinite number of nodes. 
The AAdS couterparts of the $d=5$ asymptotically flat  solutions  
are discussed in \cite{Okuyama:2002mh}. 
Although approaching asymptotically the AdS$_5$ background, 
the mass of these configurations also diverges logarithmically. 

As conjectured by several authors, this result extends to higher dimensions.
Following the approach in \cite{Okuyama:2002mh},
 we prove in Appendix A 
 the nonexistence of asymptotically flat or AdS solutions with a
finite mass in a $F(2)$ 
EYM  model given by (\ref{F2}) for any spacetime dimension $d\geq 5$
(see also the discussion in Section 3.3).
Therefore, the $d-$dimensional Schwarzschild-AdS configuration is
the only finite mass solution of the  equations (\ref{eqsF2-1})-(\ref{eqsF2-3})
and a simple version of the no hair theorem seems to hold for the
$F(2)$ (usual) EYM system in $d>4$.

We should remark that in deriving this result 
we assumed implicitely that $ m(r), \sigma(r), w(r)$
are smooth functions approaching finite values as $r \to \infty$.
A divergent asymptotic value of $m(r)$ invalidates the 
proof presented in Appendix A and also the virial argumnets in Section 3.3. 
Therefore we cannot
exclude the existence of spherically symmetric, nontrivial  solutions of 
the field equations for any $d>4$.
However, the mass of these solutions generically diverges,
although the spacetimes are still AAdS. The work of Ref. \cite{Okuyama:2002mh}
presents an extensive discussion of such AAdS
solutions for $d=5$.
Both regular and black hole solutions exist in $d=5$ for compact intervals 
of the parameter
that specifies the initial conditions at the origin or at the event horizon.
Differing from the $\Lambda=0$ case, the gauge field function $w(r)$ does not
oscillate between $1$ and $-1$ but approaches asymptotically some
finite value $w_0$, the node number being finite.
The masses of these solutions behave asymptotically as $(w_0^2-1)\log r$,
with all $w_0=\pm 1$ solutions corresponding to pure gauge configurations.

The results we found by solving numerically the equations
(\ref{eqsF2-1})-(\ref{eqsF2-3}), for $d=6,7,8$ and
several negative values of $\Lambda$ confirm that this is a generic behaviour
for $d>4$. The corresponding boundary conditions at the origin (or event
horizon) are found by taking $P=1$
in relations (\ref{r=0}), (\ref{eh}) given in Section 3.
Except for a divergent value of $m(r)$ as $r \to \infty$, according to
\begin{eqnarray}
\label{mass-div}
m(r)=M_0+\frac{\tau_1(d-3)}{4(d-5)}(w_0^2-1)^2r^{d-5},
\end{eqnarray}
the properties of these solutions are very similar to the
the more familiar $d=4$ case. For $d>5$, the asymptotic value $w_0$
of the gauge field function $w$ is also arbitrary, being fixed by the initial
parameters $w''(0)$ or $w(r_h)$ respectively, $w_0=\pm 1$ corresponding to pure
gauge configurations. Solutions for a compact interval of these parameters
were found to exist, the general structure being $\Lambda$-dependent.
Solutions with nodes in $w(r)$ were also found.
Typical $d=6$ configurations with a regular origin are presented
in Figure 1, for $\Lambda=-0.01$ and three different values of $b=-w''(0)/2$.
One can see that the mass function diverges linearly while
$\sigma(r)$ and $w(r)$ asymptotically approach some finite values.
In Figure 2 we plot the parameters $M_0$ (appearing in (\ref{mass-div}),
which in Section 4.3 we argue that it can be taken as the renormalised
mass of the solutions), $w_0$, the value $\sigma_0$ of the metric function
$\sigma$ at the origin and the minimal value $N_m$ of the metric function $N$
as a function of $b$ for a family of $d=6$ AAdS solutions with $\Lambda=-1$.
This branch ends for some finite value of $b$, where $\sigma(0) \to 0$.
Black hole solutions have been found as well, presenting the same general
features. Here also we find a continuum of solutions with arbitrary
values of $w_0$, the relevant parameter being the value of the gauge potential
at the event horizon. Similar to $d=4$, solutions appear to exist for any value
of the event horizon radius.

The drawback of the solutions in $d>5$ described above is that their ADM
masses are divergent, making their physical significance obscure
(see, however, the discussion in Appendix B).

One may hope to find a different picture by including some other matter field
in the action (\ref{action}). Such fields should interact
with the YM sector so as to compensate for the scaling behaviour of
the non Abelian fields. This excludes the dilaton field, as it can be proven
that the latter does not change this nonexistence result.
Note that $d=5$ finite mass spherically symmetric gravitating non Abelian 
solutions with a Liouville-type dilaton potential
are known to exist \cite{Chamseddine:2001hk}; however these solutions are 
asymptotically neither flat nor AdS.

%Of course, less symmetric EYM  solutions, obtained for an Ansatz with a number
%of codimensions, are very likely to exist. The simplest such example with one
%Killing symmetry $\partial/\partial z$ is given by 
%\begin{eqnarray}
%ds_d^2=F^2(r)ds_{d-1}^2+G^2(r)dz^2,
%~~~
%A=A_{\mu}dx^{\mu}+A_{d}dz
%\end{eqnarray}
%where $\mu=1,d-2$ and $A_{\mu}$ still given by (\ref{YMsph}). 
%Here $ds_{d-1}^2$ is a $d-1$ dimensional
%spherically symmetric line element.
%These solutions would describe $d-$dimensional non Abelian black strings and
%vortices (some features of such $d=5$ configurations are discussed in 
%\cite{Volkov:2001tb,Hartmann:2003kz,Brihaye:2005pz}).

In the next Section, we will remedy this problem of nonexistence by adding
higher order YM curvature terms $F(2p)$, $(p>1)$, to \re{F2}. The main role of
these terms is to alter the scaling properties of the action density in
\re{action}. But logically, such a role can be played also by altering the
gravitational part of \re{action}, through the introduction of higher order
(in the Riemann curvature) terms $R(p)$, $(p>1)$. $R(p)$ here denotes the
generalised Ricci scalar constructed from the $2p$-form antisymmetrised
$p$-fold product of Riemann tensor $2$-forms, the $p=1$ member of which gives
the Einstein-Hilbert action, the $p=2$, the Gauss--Bonnet, etc. Here we have
eschewed the possibility of employing additional $R(p)$ terms instead of,
or together with $F(2p)$ YM terms, because in the present work we exclude
the participation of fields other than gravitational and Yang-Mills. In
particular, the exclusion of the dilaton renders the usefulness of
higher order (Gauss-Bonnet) gravities trivial from a practical point of view.

To see this, consider the situation where a term
scaling as $\it{ L}^{-2p}$ is needed, i.e. that the term
\[
\ka_p\,\sqrt{-g}\,R(p)
\]
must be added to the density in \re{action}. Using the metric Ansatz
\re{metric}, and discarding purely boundary terms in the residual one
dimensional (spherically symmetric) Lagrangean, the term $\mathcal{ L}_{(p)}$ to be added
is
\be
\label{gravterm}
\mathcal{ L}_{(p)}=\ka_p\,\frac{(d-2p)!}{(d-1)!}\,\si\,r^{d-2p-2}
\left[r\frac{d}{dr}(1-N)^p+(d-2p-1)(1-N)^p\right]
\ee
in spacetime dimension $d$, which vanishes identically in dimensions up to
$d=2p$. This is because in the given dimensions the $p$-th member of the
gravitational hierarchy becomes a total divergence. But for given $d$ it is
necessary to have $2p\ge d$, whence \re{gravterm} trivialises.
The situation would be quite different if the dilaton field were included
together with the higher order gravitational terms, preventing the
$p$-gravity density becoming a total divergence. But this option is excluded
here and we opt to the exclusive use of higher order $p$-YM densities.

Yet other fields occurring in supergravities might be considered, e.g.
Kalb-Ramond, or totally antisymmetric tensor fields. But these being Abelian,
their effect would be felt only in given dimensions, or, subject to much less
stringent symmetries than the spherical. This also is not a flexible option,
so we restrict our attention to $p$-YM densities only.

%%%%%%%%%%%%%%%%%%%%%%%%%%%%%%%%%%%%%%%%%%%%%%%%%%%%%%%%%%%%%%%%%%%%%%%%%%%%%%
\section{Nonabelian hierarchies}
%%%%%%%%%%%%%%%%%%%%%%%%%%%%%%%%%%%%%%%%%%%%%%%%%%%%%%%%%%%%%%%%%%%%%%%%%%%%%%
%%%%%%%%%%%%%%%%%%%%%%%%%%%%%%%%%%%%%%%%%%%%%%%%%%%%%%
\subsection{The Lagrangean and field equations}
%%%%%%%%%%%%%%%%%%%%%%%%%%%%%%%%%%%%%%%%%%%%%%%%%%%%%%
Since finite mass spherically symmetric solutions play a central role in
AdS/CFT, it is desirable that the nonexistence result 
presented in Section {\bf 2.5} be circumvented.

A simple way to circumvent these arguments and to find 
nontrivial solutions is to to modify the matter Lagrangean
by adding higher order terms in the YM hierarchy. As noted in footnote$^1$,
these are constructed exclusively from YM curvature $2p$-forms.  
For $\Lambda=0$, asymptotically flat, finite energy solutions of this modified
EYM system are constructed in \cite{Brihaye:2002hr,Brihaye:2002jg}. 

Such terms as we propose to add are predicted by string theory, and hence
provide a link with the AdS/CFT correspondence too. But here we are guided
predominantly by symmetry considerations and do not claim to be employing
terms strictly following from superstring theory. The situation concerning
higher order YM curvature terms in the string
theory effective action is complex and as yet not fully resolved. While
YM terms up to $F^4$ arise from (the non Abelian version of) the
Born--Infeld action~\cite{Tseytlin}, it appears that this approach
does not yield all the $F^6$ terms~\cite{BRS}.
Terms of order $F^6$ and higher can also be obtained by employing the
constraints of (maximal) supersymmetry~\cite{CNT}. The results
of the various approaches are not identical.

The definition we use for superposed YM hierarchy is
\be
\label{YMhier}
{\cal L}_{m}=-\sum_{p=1}^{P}\ \frac{1}{2(2p)!}\ \tau_p~\sqrt{-g}~
{\mbox Tr\ }F(2p)^2\ ,
\ee
where $F(2p)$ is the $2p$-form $p$-fold totally antisymmetrised product
of the $SO(d)$ YM curvature $2$-form $F(2)$
\be
\label{2pformYM}
F(2p)\equiv F_{\mu_1\mu_2...\mu_{2p}}=F_{[\mu_1\mu_2}F_{\mu_3\mu_4}...
F_{\mu_{2p-1}\mu_{2p}]}\ .
\ee
Even though the $2p$-form \re{2pformYM} is dual to a total divergence,
namely the divergence of the corresponding Chern-Simons form, the density
\re{YMhier} is never a total divergence since it is the square of one.
But the $2p$-form \re{2pformYM} vanishes by (anti)symmetry for $d<2p$ so
that the upper limit in the summation in \re{YMhier} is $P=\frac{d}{2}$
for even $d$ and $P=\frac{d-1}{2}$ for odd $d$.

We define the $p$-stress tensor pertaining to each term in (\ref{YMhier}) as
\be
T_{\mu\nu}^{(p)}=
\mbox{Tr}\ F(2p)_{\mu\lambda_1\lambda_2...\lambda_{2p-1}}
F(2p)_{\nu}{}^{\lambda_1\lambda_2...\lambda_{2p-1}}
-\frac{1}{4p}g_{\mu\nu}\ \mbox{Tr}\ F(2p)_{\lambda_1\lambda_2...\lambda_{2p}}
F(2p)^{\lambda_1\lambda_2...\lambda_{2p}}\ .
\label{pstress}
\ee

For the particular spherically symmetric ansatz
considered in Section {\bf 2},
we express the reduced YM Lagrangean arising from \re{YMhier} as
\be
\label{p-split}
{\cal L}_{m}=-\sum_{p=1}^{P}\, L_{YM}^{(p)}
\ee
with $L_{YM}^{(p)}$  given by
\bea
L_{YM}^{(p)}=r^{d-2}\si\,\frac{\tau_p}{2\cdot (2p)!}\,
\frac{(d-2)!}{(d-[2p+1])!}\
W^{p-1}
\left[(2p)N\left(\frac1r\frac{dw}{dr}\right)^2+
(d-[2p+1])\,W\right]\label{ymp}
\eea
having used the shorthand notation
\be
\label{defW}
W=\left(\frac{w^2-1}{r^2}\right)^2\,.
\ee
For this general ansatz, we find the field equations
%%%%%%%%%%%% new equations %%%%%%%%%%%%%%%%
\begin{eqnarray}
\nonumber
&&m'=\sum_{p=1}^{P}\frac{\tau_p r^{d-2}}{2 (2p)!} \frac{(d-3)!}{(d-[2p+1])!}
W^{p-1}\left[(2p)N\left(\frac1r\frac{dw}{dr}\right)^2+
(d-[2p+1])\,W\right],
\\
\label{gen-eqs}
&&\frac{  \sigma'}{\sigma} =\frac{(d-3)!}{\kappa r}w'^2
\sum_{p=1}^{P_2} \frac{\tau_p W^{p-1}}{(d-[2p+1])!(2p-1)!},
\\
\nonumber
&&\sum_{p=1}^{P}\frac{d}{dr}\Big(\frac{r^{d-4}\sigma\tau_p}{(d-[2p+1])!(2p-1)!}
W^{p-1}Nw'\Big)=\sum_{p=1}^{P}
\frac{r^{d-6} \sigma \tau_p2w(w^2-1)W^{p-1}}{(d-[2p+1])!(2p-1)!}
\Big(N(p-1)W \frac{w'^2}{r^2}+\frac{d-2p-1}{2} \Big),
\end{eqnarray}
which can also be derived from the reduced action
\begin{eqnarray}
\label{red-action-gen}
S=\int dr~\sigma \Big(m'-\sum_{p=1}^{P}\frac{\tau_p r^{d-2}}{2 (2p)!} \frac{(d-3)!}{(d-[2p+1])!}
W^{p-1}\left[(2p)N\left(\frac1r\frac{dw}{dr}\right)^2+
(d-[2p+1])\,W\right]\Big).
\end{eqnarray}
For $ \tau_i=0$, $i>2$ the equations (\ref{eqsF2-1})-(\ref{eqsF2-3}) are
recovered. Note also the $\sigma$ equation decouples and can be treated
separately.

%%%%%%%%%%%%%%%%%%%%%%%%%%%%%%%%%%%%%%%%%%%%%%%%%%%%%%
\subsection{Boundary conditions}
%%%%%%%%%%%%%%%%%%%%%%%%%%%%%%%%%%%%%%%%%%%%%%%%%%%%%%

The asymptotic solutions to these equations can be systematically
constructed in both regions, near the origin (or event horizon)
and for $ r \gg 1$.
 
The corresponding expansion as $r \to 0$ is
\begin{eqnarray}
\nonumber
w(r)&=&1-b r^2+O(r^4),
\\
\label{r=0} 
m(r)&=&\Big(\sum_{p=1}^{P}\frac{\tau_p(d-3)!(4b^2)^p}{2(2p)!(d-[2p+1])!} 
\Big) r^{d-1}+O(r^{d+1}),
\\
\nonumber
\sigma(r)&=&\sigma_0+\frac{(d-3)!}{2\kappa}
\Big(\sum_{p=1}^{P}\frac{\tau_p(4b^2)^p}{(2p-1)!(d-[2p+1])!} \Big)
r^{2}+O(r^4),
\end{eqnarray}
and contains one essential parameter $b$ (the value of $\sigma(0)\equiv\si_0$
can be fixed by rescaling the time coordinate).

For black hole configurations with a regular, nonextremal event horizon at
$r=r_h$, the expression  near the event horizon is\begin{eqnarray}
\label{eh}
\nonumber
m(r)&=&m_h+m'(r_h)
(r-r_h)+O(r-r_h)^2,
\nonumber
\\
w(r)&=&w_h+w'(r_h)
(r-r_h)+O(r-r_h)^2,
\\
\nonumber
\sigma (r)&=&\sigma_h+\sigma_h'(r-r_h)+O(r-r_h)^2,
\end{eqnarray}
where
\begin{eqnarray}
\nonumber
m(r_h)=\frac{1}{2} \kappa r^{d-3}_h (1+\frac{r_h^2}{\ell^2} ),\,\
~~~~
W_h=\frac{(w_h^2-1)^2}{r_h^4},\,\,\,\,\,\,\,\,\,\,\,\,\,\,\,\
\\
\label{cond-bh}
m'(r_h)=\sum_{p=1}^{P}\frac{\tau_p r_h^{d-2}(d-3)!}{2(2p)!(d-2p-2)!}W_h^p,
~~
N'_h=\frac{d-3}{r_h}+\frac{(d-1)r_h}{\ell^2}-\frac{2m'(r_h)}{\kappa r_h^{d-3}},
\\
\nonumber
\sigma_h'=
 \frac{\sigma_h (d-3)!}{\kappa r_h}w'^2_h
\sum_{p=1}^{P}\frac{\tau_p W_h^{p-1}}{(2p-1)!(d-[2p+1])!},
~~
w'_h=\frac{1}{N'_h}\frac{w_h(w_h^2-1)}{r_h^2}
\frac{\sum_{p=1}^{P}\frac{\tau_p  W_h^{p-1}}{(2p-1)!(d-2p-2)!}}
{\sum_{p=1}^{P}\frac{\tau_p  W_h^{p-1}}{(2p-1)!(d-2p-1)!}},
\end{eqnarray}
the value of the gauge field on the event horizon being the esential parameter.
Here the obvious condition $N'(r_h)>0$ imposes some limits
on the event horizon radius as a function of $w_h$ for given
$(\tau_i,\Lambda)$.

Since the field equations are invariant under
$w \to -w$, one can take $w(0)=1$ and $w(r_h)>0$ without any loss
of generality.

For $r \gg 1$ we find for both regular and black hole solutions
\begin{eqnarray}
w(r)&=&\pm 1+\frac{w_1}{r^{d-3}}+\dots,\nonumber
\\
m(r)&=&M-\frac{\tau_1(d-3)w_1^2}{8 \ell^2}\frac{1}{r^{d-3}}+\dots,
\label{29}\\
\sigma(r)&=&1-\frac{w_1^2 (d-3)^2\tau_1}{2\kappa (d-2)}\frac{1}{r^{d-4}}+\dots.
\nonumber
\end{eqnarray}
These boundary conditions are also shared by the asymptotically flat solutions 
(with a different decay of the mass function $m(r)$, however),
$w=\pm 1$ being again the only allowed values of the gauge function as
$r \to \infty$.
Therefore, we expect to find a qualitatively similar picture in both cases.
We will find in Section {\bf 5} that the constant $M$ in the above relations
is the  ADM  mass up to a $d$-dependent factor.
However, in the discussion of numerical solutions we will refer to 
$M$ to as the mass of the solutions.

%%%%%%%%%%%%%%%%%%%%%%%%%%%%%%%%%%%%%%%%%%%%%%%%%%%%%%
\subsection{Further relations}
%%%%%%%%%%%%%%%%%%%%%%%%%%%%%%%%%%%%%%%%%%%%%%%%%%%%%%
The form (\ref{red-action-gen}) of the reduced action allow 
to derive an useful virial relation.
To this end, we  use the scaling technique  proposed  in
\cite{Heusler:1996ft, 18} for the case of spherically symmetric gravitating
systems. Let us assume the existence of a globally regular solution $ m(r), \sigma(r), w(r)$
of the field equations (\ref{gen-eqs}),
with suitable boundary conditions at the origin and at infinity. 
Then each member of the 1-parameter family
\begin{equation}
m_\lambda(r) \equiv m(\lambda r), \ 
\sigma_{\lambda} (r) \equiv \sigma (\lambda r), \ 
w_{\lambda} (r) \equiv w(\lambda r)
\end{equation}  
assumes the same boundary values at $r=0$ and $r=\infty$, and the action 
$S_{\lambda} \equiv S[m_{\lambda}, \sigma_{\lambda}, w_{\lambda}]$
must have a critical point at $\lambda=1$, $i.e.$  $[dS/d\lambda]_{\lambda=1}=0$.
Therefore we find 
the following virial
relation satisfied by the finite energy solutions of the field equations
(note that following \cite{Heusler:1996ft}, 
it is possible to write a similar relation 
for black hole configurations, also)
\begin{eqnarray}
\label{virial-gen}
\sum_{p=1}^{P}\int_0^\infty dr~\sigma 
\frac{\tau_p (d-3)!}{2(2p)!d-[2p+1])!}W^{p-1}
\Big((d-4p-1)(2pN\frac{w'^2}{r^2}+(d-2p-1)W)
\\
\nonumber
+2p\frac{w'^2}{r^2}(\frac{2(d-3)m}{\kappa r^{d-3}}+\frac{2r^2}{\ell^2})\Big)=0.
\end{eqnarray}
For $p=1$, $i.e.$ a $F^2$-theory, the above relation reads
\begin{eqnarray} 
\label{virial1}
\int_{0}^{\infty}  dr ~\sigma~r^{d-4}
\Big( (d-5)(N w'^2+\frac{(w^2-1)^2}{2r^2})
+w'^2(\frac{2m}{\kappa r^{d-3}}(d-3)+\frac{2r^2}{\ell^2}) \Big)=0,
\end{eqnarray}
which clearly shows that no nontrivial gravitating solution with finite mass exists for $d>4$,
since all terms in the integrant are strictly positive quantities.

Therefore it becomes obvious that new terms in the YM-hierarchy
should be introduced as the spacetime dimension increases.
For a given $d$, the relation $P>[(d+1)/4]$ should be satisfied.
As it happens, to go to $5\leq d< 9$ it is necessary to include at least the
second member of the YM hierarchy to provide the requisite scaling (similarly
for $9\leq d< 13$ it is necessary to include the thrid  member of the YM
hierarchy). In practice we add only the lowest order such term necessary.

%%%%%%%%%%%%%%%%%%%%%%%%%% Hawking temperature 
 %%%%%%%%%%%%%%%%%%%%%%%%%%%%%%%%%%%
We mention here also the Hawking temperature  expression of the 
 black hole solutions. 
For the line element (\ref{metric}), if we treat $t$ as complex, 
then its imaginary part is a coordinate for a nonsingular 
Euclidean submanifold iff it is periodic with period 
\begin{eqnarray}\label{period}
\beta=\frac{4 \pi }{N'(r_h)\sigma(r_h)}.
\end{eqnarray}
Then continuous Euclidean Green functions must have this period, so by standard
arguments the Hawking temperature is (with $k_B=\hbar=1$)
\begin{eqnarray}
\label{Temp}
T_H=\frac{\sigma_h}{4 \pi r_h}
\Big( d-3-\frac{2m'(r_h)}{\kappa r_h^{d-3}}+(d-1)\frac{r_h^2}{\ell^2}\Big) 
\leq \frac{\sigma_h}{4 \pi r_h}
\Big(d-3+(d-1)\frac{r_h^2}{\ell^2}\Big). 
\end{eqnarray} 
Thus the Hawking temperature of such systems appears to be suppressed relative
to that of a vacuum black hole of equal  horizon area.

%%%%%%%%%%%%%%%%%%%%%%%%%% rescaling %%%%%%%%%%%%%%%%%%%%%%%%%%%%%%%%%%%
In the presence of higher oder terms in the YM action,
dimensionless quantities are obtained by rescaling 
\begin{eqnarray}
 r \to (\tau_2/\tau_1)^{1/4},~~\Lambda \to (\tau_1/\tau_2)^{1/2},
~~m(r) \to m(r)\kappa(\tau_1/\tau_2)^{(d-3)/4},
\end{eqnarray}
This reveals the existence of 
one fundamental parameter which gives the strength of
the gravitational interaction
\begin{eqnarray}
 \alpha^2=\frac{\tau_1^{3/2}}{\kappa \tau_2^{1/2}},
\end{eqnarray}
and $P-2$ independent coupling constants
\begin{eqnarray}
 \beta^2_{k-2}=\frac{\tau_k}{\tau_1}\left(\frac{\tau_1}{\tau_2}\right)^{k-1},
\end{eqnarray}
with $k=3,P-2$ (i.e. no such constants appear in a $p=2$ system).

For the $F(2)+F(4)$ systems in $d=6,7,8$ considered in \cite{Brihaye:2002hr},
there exist
gravity decoupling solutions at $\al=0$, from which the gravitating solutions
branch out to a maximum value of $\al_{\rm max}$ and then decrease. The
second limit of $\alpha \to 0$ also exists. For the same system in $d=5$
studied in \cite{Brihaye:2002jg} on the other hand, there exists no gravity
decoupling limit but nonetheless the solution branches out from $\al=0$
by employing a scaling procedure \cite{Breitenlohner:2005}, and again
$\al$ increases to a value $\al_{\rm max}$ and decreases. But in this case it
does not reach the $\alpha \to 0$ limit. Rather it stops at a new critical
point $\al_{\rm c}$ around which it oscillates \cite{Brihaye:2002jg}, which is
a new type of critical point identified in \cite{Breitenlohner:2005} and named
a 'conical' fixed point.

%%%%%%%%%%%%%%%%%%%%%%%%%%%%%%%%%%%%%%%%%%%%%%%%%%%%%%%%%%%%%%%%%%%%%%%%%%%%%%
%%%%%%                      NUMERICAL SOLUTION
%%%%%%%%%%%%%%%%%%%%%%%%%%%%%%%%%%%%%%%%%%%%%%%%%%%%%%%%%%%%%%%%%%%%%%%%%%%%%%
\subsection{Numerical solutions}
%%%%%%%%%%%%%%%%%%%%%%%%%%%%%%%%%%%%%%%%%%%%%%%%%%%%%%%%%%%%%%%%%%%%%%%%%%%%%%
In the present work, we restrict our attention to the
simplest nontrivial cases with
only the two terms $p=1$ and $2$ in the
YM hierarchy. 
However, we have obtained some numerical results also for a $P=3$
hierarchy with $\beta_1 = 1$,
which will be briefly mentioned. 
For $d=5$, the solutions we found have some special features which
will be discussed separately. 

Both regular and black hole solutions of the EYM-hierarchy equations
appear to exist for any value of $\Lambda$.
Given ($\alpha, d,\Lambda$), 
AAdS solutions may exist for discrete set of shooting parameters
 $b$ and $w_h$ respectively.
We follow the usual approach and, by using a standard ordinary  differential  
equation solver, we evaluate  the  initial  conditions  at  $r=10^{-5}$
(or $r_h+10^{-5})$ for  global  tolerance  
$10^{-14}$, adjusting  for shooting parameters and  integrating 
towards  $r\to\infty$, and looking for AAdS solutions with a finite mass.

Similar to the $p=1,~d=4$ asymptotically flat case, 
it can be proven that for all AAdS (or $\Lambda=0$)
solutions, $w(r)$ is confined within the strip $|w(r)|<1$.
This can proven as follows: we suppose the existence of solutions with $w(r)>1$
for some interval of $r$.
Therefore $w$ must develop a maximum for some $r_0$, 
$w'(r_0)=0$ and $w(r_0)>1$ with $w''(r_0)<0$.
However, the equations (\ref{gen-eqs})
imply that in the region $w>1$ the only extremum can be a minimum.
Therefore the condition $|w(r)|<1$ is always fullfiled.
As a general feature, 
all solutions discussed in the rest of this section present only one 
node in the gauge function $w(r)$. 
Similar to the $\Lambda=0$ case, we could not find multinode solutions
\footnote{Multinode solutions exist if
the lowest order YM term is $F(2p)$ with $p\ge 2$, in the appropriate
dimensions \cite{Breitenlohner:2005} $d$.}.

The absence of multinode solutions in this $F(2)+F(4)$ model with $\Lambda=0$,
in the relevant dimensions $5\le d\le 8$ is analytically explained in
\cite{Breitenlohner:2005}. We expect that the relevant fixed point analysis
yields qualitatively similar results also for $|\Lambda|>0$. This is borne
out by our numerical results.

For any regular solution, the metric functions $m(r)$ and $\sigma(r)$ 
always increase monotonically
with growing $r$ from 
$m(0)=0$ and $\sigma(0)=\sigma_0$ at the origin to $m(\infty)=M$ and
$\sigma(\infty)=1$, respectively.
The gauge function always interpolates between $w(0)=1$ and $w(\infty)=-1$ 
without any local extrema.
For black hole configurations, the behaviour of the functions $m(r),\sigma(r)$
and $w(r)$ is similar to that for regular solutions.
The gauge potential $w(r)$ starts from some finite value $0<w(r_h)<1$ at the
horizon and monotonically approaches $-1$ at infinity.
The metric functions $m(r)$ and $\sigma(r)$ 
increase also monotonically with $r$.
In the asymptotic region, the geometry corresponds to 
a Schwarzschild-AdS solution.
However, although these solutions are static and have 
vanishing YM charges $(w^2(\infty)=1)$,
they are different from the Schwarzschild-AdS black hole,
and therefore are not fully characterised by the mass-parameter $M$.

%%%%%%%%%%%%%%%%%%%%%%%%%%%%%%%%%%%%%%%%%%%%%%%%%%%%%%%%%%%
\subsubsection{Regular solutions $d=5$}
%%%%%%%%%%%%%%%%%%%%%%%%%%%%%%%%%%%%%%%%%%%%%%%%%%%%%%%%%%%
As $\alpha^2 \rightarrow 0$, the YM equations present a nontrivial,
finite energy solution in a fixed AdS background.
This nongravitating configuration approaches in the $\Lambda=0$ limit the
YM instanton in four dimensional flat space \cite{Belavin:1975fg}.

When $\alpha^2$ increases, this solution gets deformed by gravity 
and the mass $M$ decreases. At the same time,
both the value $\sigma(0)$ and the minimal value $N_m$ of the function $N(r)$
decrease, as indicated in Figure 3.
This branch of solutions exists up to a maximal
value $\alpha_{\rm max}$ of the parameter $\alpha$,
which is smaller than the corresponding value in
the asymptotically flat case~\cite{Brihaye:2002jg}.
For example, we find numerically
$\alpha_{\rm max}^2 \approx 0.3445$ for $\Lambda=-0.2$ while the
corresponding value for $\Lambda=-0.01$ is $\alpha_{\rm max}^2 \approx 0.5322$.
(without a cosmological term, this branch extends up to
 $\alpha_{\rm max}^2 \approx 0.5648$.)

Similar to the EYM theory with $\Lambda=0$ \cite{Brihaye:2002jg}, 
we found always another branch of solutions
on the interval $\alpha^2 \in [\alpha_{cr(1)}^2 , \alpha_{\rm max}^2]$
with $\alpha_{cr(1)}^2 $ depending again on the value of $\Lambda$ 
(e.g. $\alpha_{cr(1)}^2  \approx 0.2876$ for $\Lambda=-0.2$).
On this second branch of solutions, both $\sigma(0)$ and $N_m$ continue
to decrease but stay finite.
 However, a third branch of solutions exists
for $\alpha^2 \in [\alpha_{cr(1)}^2 , \alpha_{cr(2)}^2]$ ,
on which the two quantities decrease
further.
A fourth branch of solutions has also been
found, with a corresponding $\alpha_{cr(3)}^2$ close to
$\alpha_{cr(2)}^2$. Further branches of solutions, exhibiting more
oscillations very likely
exist but their study is a difficult numerical problem.
Along this succession of branches, the main observation
is that the value $\sigma(0)$ decreases much faster than that of $N_m$
as illustrated in Figure 3. Also, the mass parameters do not increase
significantly along these secondary branches.
However, the shooting parameter $b$ increases to very large values.
 The pattern strongly suggests that after a
finite (or more likely infinite) number of oscilations of $\sigma(0)$,
the solution terminates into a singular solution with $\sigma(0)=0$
and a finite value of $N(0)$.

This is the behaviour observed in \cite{Brihaye:2002jg} for the EYM theory
with $\Lambda=0$. The inclusion of a negative cosmological constant does not
seem to qualitatively change the properties of the system, but leads to
different values of the critical parameters. As in the $\Lambda=0$ case,
the dominant term at the gravity decoupling limit $\al\to 0$ is the $F(2)$
term, the energy being given by the
action of the usual instanton \cite{Belavin:1975fg}, while as $\al\to\al_{cr}$
the dominant term is $F(4)$. The mechanism for this effect is explained in
\cite{Breitenlohner:2005} (for $\Lambda=0$) and is supported by our
numerical results here. The typical $d=5$ globally regular 
solutions look very similar to the $d=6,8$ profiles presented in Figures 4, 5.

%%%%%%%%%%%%%%%%%%%%%%%%%%%%%%%%%%%%%%%%%%%%%%%%%%%%%%%%%%%
\subsubsection{Regular solutions $d>5$}
%%%%%%%%%%%%%%%%%%%%%%%%%%%%%%%%%%%%%%%%%%%%%%%%%%%%%%%%%%
The solutions in this case resemble again the $\Lambda=0$ situation.
In the presence of
suitable higher order term in the hierarchy, 
the YM equations admit finite energy solutions in a fixed AdS$_d$
A $p=1$, AdS$_4$ exact solution was found in \cite{Boutaleb-Joutei:1979va}, 
its higher order generalisation (specifically for $p=2$, AdS$_8$) for $d>4$
being discussed in \cite{Brihaye:2000cz}.
According to the standard arguments, this AdS soliton can be generalised
in the presence 
of gravity, provided that the dimensionless coupling constant $\alpha$
is small enough.
Therefore the gravitating solutions exist up to a maximal value
$\al_{\rm max}$ of the  gravitational coupling constant.
% and persist for smaller $\al$, down to $\al=0$. 
%The numerical
%process however does not converge for some values of $\al$ depending
%on $d$, say $\al_d$. 
This value $\alpha_{\rm max}$ for a given $d$ depends on
the value of $\Lambda$.
For example in $d=6$, $\alpha_{\rm max}(\Lambda=0)=0.12675$; 
$\alpha_{\rm max}(\Lambda=-1)=0.070422$; 
in 8 dimensions we find $\alpha_{\rm max}(\Lambda=0)=0.002193$
while $\alpha_{\rm max}(\Lambda=-5)=6.69\times 10^{-4} $.

When $\alpha$ increases, the  mass of the gravitating solutions decreases
while the function $N(r)$ develops a local minimum
$N_m$ which becomes deeper while gravity becomes stronger
and the value $\sigma(0)$ decreases from one.
At the same time, the value of the shooting parameter $b$ increases with
$\alpha$. Our numerical analysis for $d\leq 10$ indicates that a second branch
of regular solutions always exists, starting at $\alpha_{\rm max}$.
Along this  second branch the values $\sigma(0)$
and $N_m$ decrease monotonically with $\alpha$, while $b$ and $M$ still
increase. The mass of a second branch solution 
is always larger than the corresponding mass (for the same value of
$\alpha$) on the first branch.
For $d>6$, the numerical analysis suggests that this second branch persists
up to $\alpha^2 \simeq 0$ and that in this limit $\sigma(0)$ approaches a
very small value. As far as our numerical analysis indicates, the value $N_m$
tends to a finite value in this limit so that there occurs no horizon.
Therefore two regular solutions seems always to exist 
for any $\alpha<\alpha_d$.

The case $d=6$ is special, since the numerical procedure fails to
give reliable results for second branch solutions, starting with some
 $\alpha_d$, whose value is $\Lambda$-dependent (for example we found
$\alpha_d(\Lambda=-1)=0.0435$ while $\alpha_d(\Lambda=0)=0.0573$).
The quantity $\sigma(0)$ reaches a very small value as
$\alpha \rightarrow \alpha_d$. The minimal value of $N(r)$ remains finite
so that no horizon is approached. 
We expect that a different parametrisation of the metric and variables would
allow us to continue this second branch to $\alpha \to 0$ in this case, too.

The behaviours just described qualitatively duplicate those of the $\Lambda=0$
case \cite{Brihaye:2002hr}, and are analysed in \cite{Breitenlohner:2005}.
Likewise, the solutions are dominated by the $F(2)$ terms in the gravity
decoupling limit $\al\to 0$, while at the other end (on the second mass-branch)
it is the $F(4)$ terms that dominate.

Typical $d=6,8$ solutions are presented in Figures 4, 5 respectively.
In Figures 5, 6 and 7 we plot some relevant quantities for $d=6,7,8$
and several values of $\Lambda$.
One can see that the qualitative behaviour of the  
functions $m,~\sigma,~w$ does not change by changing the 
value of $\Lambda$. 

The results we found by including the $p=3$ term in the YM-hierarchy for
$d=9,10$ follows the same pattern. 
Although the picture gets more complicated by the existence of one 
more coupling parameter, two branches of solution are always found to exist.
We noticed also the existence of a maximal value of $\alpha$ which is
$(\Lambda, \beta_1)$ dependent.

%%%%%%%%%%%%%%%%%%%%%%%%%%%%%%%%%%%%%%%%%%%%%%%%%%%%%%%%%%%
\subsubsection{Black hole solutions $d=5$}
%%%%%%%%%%%%%%%%%%%%%%%%%%%%%%%%%%%%%%%%%%%%%%%%%%%%%%%%%%

According to the standard arguments, one can expect
to find black hole generalisations for any regular configurations, at
least for small values of the horizon radius $r_h$.
For completeness we discuss here the basic features 
of the AAdS black hole solutions.

Again, the case $d=5$ is special.
The properties of these AAdS solutions 
are rather similar to the five dimensional
asymptotically flat black hole configurations discussed in
\cite{Brihaye:2002jg}.
First, black hole solutions seem to exist for all values of $\alpha$
for which regular solutions could be constructed. Also, solutions exist only
for a limited region of the $(r_h,\alpha)$ space.

The typical behaviour of solutions as a function of $r_h$ is presented in
Figure 9, for a small value of $\alpha$ compared to the maximal value
$\alpha_{\rm max}$ of the regular solutions, in which case 
we notice the existence of only one $r_h=0$ regular configuration.
Starting from this regular solution and increasing the event horizon radius,
we find a first branch of solutions which extends to a maximal value
$r_{h(max)}$. As seen in Figures 9, 10, the value of $r_{h(max)}$
depends on $\Lambda,~\alpha$. The Hawking temperature decreases on this branch,
while the mass parameter increases; however,
the variation of mass and $\sigma(r_h)$ is relatively small. 
Extending backwards in $r_h$, we find a second branch of
solutions for $r_h < r_{h(max)}$.
This second branch stops at some critical value $r_{h(cr)}$,
where the numerical iteration fails to converge.
The value of $\sigma(r_h)$ on this branch
decreases drastically, as shown in Figure 9.
Also, the Hawking temperature after initially increasing,
strongly decreases for values near $r_{h(cr)}$, 
approaching a very  small value, while
the increase of the total mass is still very small.
Similar to the $\Lambda=0$ case~\cite{Brihaye:2002jg}, 
higher branches of solutions
on which the value $\sigma(r_h)$ continues to decrease further to zero
are likely to exist. However,
the extension of these branches in $r_h$ will be very small,
which makes their study difficult.
An approach to this problem with different parametrisation 
appears to be necessary.

However we find, that the global picture is changed by considering large
enough values of $\alpha$. In this case, more than one regular configuration
exists for a given value of $\alpha$.
This situation is illustrated in Figure 10, for solutions with $\alpha^2=0.5$.
Two regular solutions exist for this particular value of $\alpha$ and
we find two black hole branches connecting these $r_h=0$
configurations. Again, the mass of the second branch solutions is always larger
than the corresponding mass on the first branch.

Preliminary numerical results indicate an even more complicated picture for
solutions with $\alpha$ near $\alpha_{\rm max}$.
In this case the configurations combine features of both types of solutions
discussed above. Several branches of black hole solutions are found
for the same $\alpha$. These branches start from regular configurations and are
possibly disconnected.

%%%%%%%%%%%%%%%%%%%%%%%%%%%%%%%%%%%%%%%%%%%%%%%%%%%%%%%%%%%
\subsubsection{Black hole solutions $d>5$}
%%%%%%%%%%%%%%%%%%%%%%%%%%%%%%%%%%%%%%%%%%%%%%%%%%%%%%%%%%
Although predicted in \cite{Brihaye:2002jg},
no discussion of the $\Lambda=0$, $d>5$ black holes is presented in literature.

Again, black hole counterparts appear to exist for any regular solution.
However, solutions with the right asymptotics are found
for a limited region of the $(r_h,\alpha)$ space only.
We plot in Figures 11-13  some results we found for $d=6,~7$ and $d=8$.
Starting for a given $\alpha_0<\alpha_{\rm max}$ from a $r_h=0$ first branch
regular solution, we found the existence of a branch of black hole solutions
extending up to a maximal value of the event horizon radius
$r_h=r_h^{\rm max}$.
When $r_h$ increases, the  mass and the Hawking temperature increase
while the value $\sigma(r_h)$ decreases from its value at the regular
solution. A second branch of black hole solutions seems to appear always at
$r_h^{\rm max}$, extending  backwards to a zero event horizon radius.
This limiting solution corresponds to the second branch of the
regular solution at this value of $\alpha=\alpha_0$. Like in those $d=5$
cases where there are two regular solutions for a given $\al$, say $\al_0$,
here too these two regular solutions are the $r_h\to 0$ limits of the
corresponding black hole "loop", connecting the two regular solutions. 
Along this second branch the values $\sigma(r_h)$  
of the metric function $\sigma$ on the event horizon 
decrease monotonically with $r_h$, while
the Hawking temperature strongly increases. 
The mass of the solution
of the second branch is larger than the corresponding one on the
first branch, as illustrated by Figures 10-12.
It is interesting to note here from the point of view of numerics, that
a black hole loop corresponding to the two $r_h\to 0$ limits of two regular
solutions, say at $\al_0$, can be constructed numerically even when the
numerical process for the higher mass branch of the regular solutions for
this $\al_0$ runs into difficulties.

The introduction of a negative cosmological constant 
does not appear to change this picture qualitatively.
However, we notice a smaller value of $r_h^{\rm max}$ with increasing
$|\Lambda|$ and a larger value of $M$ for the same $\alpha$.

In Figure~14, we present the profiles of
the metric functions $N$ and $\sigma$ and gauge function $w(r)$
for the same values of
$\alpha^2,~\Lambda$ on the first and second branch for some $d=6$ solutions.
The dependence of these functions on the value of $\Lambda$ is
illustrated in Figure 15 for several $d=7$ black hole solutions.

%%%%%%%%%%%%%%%%%%%%%%%%%%%%%%%%%%%%%%%%%%%%%%%%%%%%%%%%%%%%%%%%%%%%%%%%%%%%%%
%%%%%%%%%%%%%%%%%%%%%%%%%%%%%%%%%%%%%%%%%%%%%%%%%%%%%%%%%%%%%%%%%%%%%%%%%%%%%%
%%%%%%               ADS/CFT
%%%%%%%%%%%%%%%%%%%%%%%%%%%%%%%%%%%%%%%%%%%%%%%%%%%%%%%%%%%%%%%%%%%%%%%%%%%%%%
\section{A  computation of mass and action}
%%%%%%%%%%%%%%%%%%%%%%%%%%%%%%%%%%%%%%%%%%%%%%%%%%%%%%%%%%%%%%%%%%%%%%%%%%%%%%
\subsection{The counterterm method}
It is well known that the generalisation of Komar's formula for
AAdS spacetimes is not straightforward and 
requires the further subtraction of a background configuration in order 
to render a finite result for the mass. This problem 
was addressed for the first time in the eighties,
with different approaches (see for instance 
Ref. \cite{Abbott:1982ff, Henneaux:1985tv}).
Another formalism to define conserved charges in AAdS 
spacetimes was proposed in \cite{Ashtekar}
and uses conformal techniques to construct conserved quantities
yielding the results obtained by Hamiltonian methods.
Other more recent approaches to the same problem are presented in 
Ref. \cite{Aros:1999id}.
 
As expected, these different methods yield the same total mass for 
the spherically symmetric AAdS configurations considered in Section 3 
\begin{equation}
\label{madm}
M_{ADM}=\frac{(d-2)\Omega_{d-2}}{8 \pi G}M,
\end{equation}
where $\Omega_{d-2}=2 \pi^{(d-1)/2}/\Gamma((d-1)/2)$ is the 
area of a unit $(d-2)$-dimensional sphere, and $M$ is defined in the
second member of \re{29}.

A procedure leading (for odd dimensions) to a different
result has been proposed by  Balasubramanian
and Kraus \cite{Balasubramanian:1999re}.
This technique was inspired by AdS/CFT correspondence and consists 
in adding suitable counterterms $I_{ct}$
to the action of the theory in order to ensure the finiteness of the boundary
stress tensor derived by the quasilocal energy definition \cite{Brown:1993br}. 
These counterterms are built up with
curvature invariants of a boundary $\partial \cal{M}$ 
(which is sent to infinity after the integration)
and thus obviously they do not alter the bulk equations of motion.
Unlike background subtraction, the counterterm approach  
does no require the identification of a reference spacetime.
Given the potential relevance of the EYM solutions in an AdS/CFT context, 
we present here a computation of the boundary stress tensor and of the total
mass, by using the counterterm prescription.

The following counterterms are sufficient to cancel divergences 
in a pure gravity theory for 
$d\leq 9$, with several exceptions (see e. g. \cite{Taylor-Robinson:2000xw})
\begin{eqnarray}
\label{ct}
I_{\rm ct}=-\frac{1}{8 \pi G} \int_{\partial {\cal M}}d^{d-1}x\sqrt{-h}
\Biggl[
\frac{d-2}{\ell}+\frac{\ell \Theta(d-4)}{2(d-3)}{\cal{R}}
+\frac{\ell^3 \Theta(d-6)}{2(d-5)(d-3)^2}
\Big({\cal R}_{AB}{\cal R}^{AB}-\frac{d-1}{4(d-2)}{\cal R}^2 \Big)
\\
\nonumber
+\frac{\ell ^{5}{\sf \Theta }\left( d-8\right) }{(d-2)^{3}(d-4)(d-6)}%
\left( \frac{3d+2}{4(d-1)}{\cal RR}^{AB}{\cal R}_{AB}-
\frac{d(d+2)}{16(d-1)^{2}%
}{\cal R}^{3}\right.  
\\
\nonumber 
-2{\cal R}^{AB}{\cal R}^{CD}{\cal R}_{ABCD}-
\frac{d}{4(d-1)}\nabla _{A}{\cal R}\nabla ^{A}{\cal R}+\nabla ^{C}%
{\cal R}^{AB}\nabla _{C}{\cal R}_{AB} \Big)
\Bigg]\ .
\end{eqnarray}
Here ${\cal R}_{ABCD},~{\cal R}_{AB}$ are the Riemann and Ricci tensors, 
${\cal R}$ is the Ricci scalar for the boundary metric $h$ and $\Theta(x)$ 
is the step function, which is equal to 1 for $x\geq 0$ and zero otherwise; 
$A,B,\ldots $ indicate the intrinsic coordinates of the boundary.

Using these counterterms one can
construct a divergenceless boundary stress tensor, which is given by the 
variation of the total action at the boundary with respect to $h_{AB}$.
Its explicit expression, restricting for simplicity to $d<7$, is
\begin{eqnarray}
\label{TAB}
T_{AB}=\frac{1}{8\pi G}(K_{AB}-Kh_{AB}-\frac{d-2}{\ell}h_{AB}
+\frac{\ell}{d-3}E_{AB}),
\end{eqnarray}
where
%\be
%\label{K}
$K_{AB}=-\frac{1}{2}(\nabla_An_B+\nabla_B n_A)$
%\ee
is the extrinsic curvature defined in terms of the normal $n_A$ to the
boundary, $K$ is its trace, and $E_{AB}$ is the Einstein tensor of the
intrinsic metric $h_{AB}$.
(The corresponding form of \re{TAB} for $d=7,8$ is given e.g. in
\cite{Das:2000cu}).

The result we find in this way for $T_{AB}$ is given by
\begin{eqnarray}
T_A^B=\frac{1}{8 \pi G \ell}\Big(M+\sum_p (-1)^p\ell^{d-3}\
\frac{\Gamma (\frac{2p-1%
}{2})}{2\sqrt{\pi }\Gamma (p+1)}\delta _{2p,d-1} \Big)
\Big((d-1)u_A u^B+\delta_A^B\Big)
\frac{1}{r^{d-1}}+O(\frac{1}{r^{d}}),
\end{eqnarray}
where $u_A=\delta_A^t$ and $p$ is an integer.
We can use this approach to assign a mass-energy to an AAdS geometry
by writing the boundary metric in an ADM form
\begin{eqnarray}
\label{b-AdS}
h_{AB}dx^{A} dx^{B}=-N_{\Sigma}^2dt^2
+\sigma_{ab}(dx^a+N_{\sigma}^a dt) (dx^b+N_{\sigma}^b dt)
\end{eqnarray}
and the definition of the energy in this context is
\begin{eqnarray}
\label{mass0}
E=\int_{\partial \Sigma}d^{d-1}x\sqrt{\sigma}N_{\Sigma}\epsilon.
\end{eqnarray}
Here $\epsilon=u^{\mu}u^{\nu}T_{\mu \nu}$ is the proper energy density while
$u^{\mu}$ is a timelike unit normal to $\Sigma$.

If there are matter fields on $\cal{M}$, additional 
counterterms may be needed to regulate the action. 
This is the case of $F(2)$ theory, discussed in  Appendix B.
The couinterterm action depends in this case not only on the boundary metric but
also on the boundary value of the gauge field.

However, we find that for $P>1$ EYM solutions with $\Lambda<0$ in $d=5,6,7,8$
dimensions, the prescription  (\ref{ct}) removes all divergences. The
use of higher order terms in the YM curvature, namely $F(2p)$ forms with
$p>1$ introduced in Section {\bf 3}, results in this regularising of the
masses. The crucial point here is that these solutions approaches
asymptotically a Schwarzschild-AdS background,
and the YM asymptotic parameter $w_1$ appears only
in the next to leading order of the $T_{AB}$ expression.

The mass-energy of solutions computed in this way is
\begin{eqnarray}
\label{mass}
E=\frac{(d-2)\Omega_{d-2}}{8 \pi G} M+E_0 
\end{eqnarray}
where, for $3<d<9$
\begin{eqnarray}
\label{mass-ct}
E_0=\frac{(d-2)\Omega_{d-2}}{16 \pi G} 
\Big(\frac{3}{4}\ell^2\delta_{5,d}-\frac{5}{8}\ell^4 \delta_{d,7}\Big).
\end{eqnarray}
The additional term $E_0$ appearing in $E$ for $d=5,(7)$
is the mass of pure global $AdS_{5,(7)}$ and is usually interpreted as the 
energy dual to the Casimir energy  of the CFT defined on a four (six)
dimensional Einstein universe \cite{Balasubramanian:1999re}.

The metric restricted to the boundary $h_{AB}$ diverges due to an infinite
conformal factor $r^2/\ell^2$. The background metric upon which the dual field
theory resides is
\begin{eqnarray}
\gamma_{AB}=\lim_{r \rightarrow \infty} \frac{\ell^2}{r^2}h_{AB}.
\end{eqnarray}
For the asymptotically AdS$_d$ solutions considered here, the $(d-1$) 
dimensional boundary is the Einstein universe,
with the line element
\begin{eqnarray}
\label{b-metric}
\gamma_{AB}dx^A dx^B=-dt^2+\ell^2d\Omega^2_{d-2}.
\end{eqnarray}

In light of the AdS/CFT correspondence, Balasubramanian 
and Kraus have interpreted 
Eq. (\ref{TAB}) as $<\tau^{AB}>=\frac{2}{\sqrt{-\gamma}} 
\frac{\delta S_{eff}}{\delta \gamma_{AB}}$,
where $<\tau^{AB}>$ is the expectation value of the CFT stress tensor.
Then, the divergences which appear are simply the standard ultraviolet
divergences of a quantum field theory and we can cancel them by adding local
counterterms to the action. Corresponding to the boundary metric
(\ref{b-metric}), the stress-energy tensor $\tau_{AB}$
for the dual theory can be calculated using the following relation
\cite{Myers:1999qn}
\begin{eqnarray}
\label{r1}
\sqrt{-\gamma}\gamma^{AB}<\tau_{BC}>=
\lim_{r \rightarrow \infty} \sqrt{-h} h^{AB}T_{BC}.
\end{eqnarray}

%%%%%%%%%%%%%%%%%%%%%%%%%%%%%%%%%%%%%%%%%%%%%%%%%%%%%%%%%%%%%%%%%%%%%%%%%%%%%%
\subsection{Action and entropy}
%%%%%%%%%%%%%%%%%%%%%%%%%%%%%%%%%%%%%%%%%%%%%%%%%%%%%%%%%%%%%%%%%%%%%%%%%%%%%%
The above approach can be used to compute the Euclidean action
and to prove in a  rigorous way 
that the entropy of the EYM black hole solutions is 
one quarter of the event horizon area, as expected.
Here we start by constructing the path integral \cite{Gibbons:1976ue}
\begin{eqnarray}
\label{Z1}
Z=\int D[g]D[\Psi]e ^{-iI[g,\Psi]}
\end{eqnarray}
by integrating over all metrics and matter fields between  some given initial
and final hypersurfaces, $\Psi$ corresponding here to the SU(2) potentials.
By analytically continuing the time coordinate $t \to i\tau$,
the path integral formally converges, and in the leading order one obtains
\begin{eqnarray}
\label{Z2}
Z \simeq e^{-I_{cl}}
\end{eqnarray}
where $I_{cl}$ is the classical action evaluated on the equations of motion
of the gravity/matter system.
The physical interpretation of this formalism is that
the class of regular stationary metrics forms an ensemble of
thermodynamic systems at equilibrium temperature
$T$ (see e.g. \cite{Mann:2002fg}).
$Z$ has  the interpretation of partition function and we can
define the free energy of the system
$F=-\beta^{-1} \log Z$.
Therefore
\begin{eqnarray}
\label{i1}
\log Z=-\beta F=S-\beta E,
\end{eqnarray}
or
\begin{eqnarray}
\label{i2}
S=\beta E-I_{cl},
\end{eqnarray}
straightforwardly follows.

To compute $I_{cl}$, we make use of the Einstein equations, 
replacing the $R-2\Lambda$ volume term with
$2R_t^t-16\pi G T_t^t$.
For our purely magnetic ansatz, the term $T_t^t$  exactly cancels the matter
field Lagrangean in the bulk action.
The divergent contribution given by the surface integral term at infinity in 
$R_t^t$ is also canceled by 
$I_{\rm{surface}}+I_{ct}$ and for $3<d<9$ we arrive at the simple finite
expression
\begin{eqnarray}
\label{itot}
I_{cl}=\frac{\beta\Omega_{d-2}}{8 \pi G}
\left( 
M-\frac{r_h^{d-2}}{\ell^2}+\frac{3}{8}\ell^2\delta_{d,5}
-\frac{5}{16}\ell^4\delta_{d,7}
\right).
\end{eqnarray}
Replacing $I_{cl}$ now in (\ref{i2}) (where $E$ is the mass-energy computed in
Section {\bf 4.1}), we find
\begin{eqnarray}
\label{i3}
S= \frac{1}{4}\Omega_{d-2} r_h^{d-2},
\end{eqnarray}
which is one quarter of the event horizon area, as expected.

>From the AdS/CFT correspondence, we expect the nonabelian hairy black holes to
be described by some thermal states
in a dual theory formulated in a Einstein universe background.
The spherically  symmetric solitons will correspond to zero-temperature states 
in the same theory. 
The existence of these hairy configurations suggest that there should be some
observables in the dual CFT that encode the hair information.

%%%%%%%%%%%%%%%%%%%%%%%%%%%%%%%%%%%%%%%%%%%%%%%%%%%%%%%%%%%%%%%%%%%%%%%%%%%%%%
\section{Conclusions}
%%%%%%%%%%%%%%%%%%%%%%%%%%%%%%%%%%%%%%%%%%%%%%%%%%%%%%%%%%%%%%%%%%%%%%%%%%%%%%
Motivated by recent results in EYM theory in four dimensional
AAdS spacetime, we studied
higher dimensional spherically symmetric solutions with non Abelian fields
in the presence of a negative cosmological constant. 
Since the mass-energy of the AAdS configurations plays a central role in its application
to the AdS/CFT correspondence, we emphasised this aspect of the solutions we
found.  
The mass-energy of the usual EYM solutions  (both asymptotically flat and AAdS),
defined according to the standard prescription, always diverges in spacetime dimensions $d>4$. 
One of the tasks performed in this work
was a demonstration of this fact in the AAdS case. 
The  main properties of these higher dimensional $F(2)$ solutions resemble the
$d=4$ case, a continuum of solutions  with arbitrary asymptotic values of the
gauge function $w(r)$ being found.

Then we focused on two
possible approaches of dealing with the divergences of the mass and 
action. One
of these involved the regularisation of the mass-energy using a 
counterterm
mechanisms. In this approach,  it turns out that  the counterterm 
action
depends not only on the boundary metric, but
also on the boundary values of the gauge fields.  However, the
masses of the solutions defined in this way may take negative values,
leading us to the second approach.  

The other, which formed the main thrust of the work, was to
augment the action density of the system with higher order curvature terms,
consisting of $2p$-form curvatures $F(2p)$. These terms were added to the
usual YM system constructed from $F(2)$. It resulted in EYM solutions
supporting finite mass-energy in all spacetime dimensions $5\le d\le 2p$.

Concerning the construction of regular finite energy classical AAdS solutions
in higher dimensions, we restricted ourselves to systems consisting exclusively
of gravitational and non Abelian gauge fields. The salient features of the
resulting solutions are captured in this framework, the addition of other
(string theory inspired) matter terms being deferred to later work.
The asymptotically flat versions of the higher-$p$ EYM systems having been
studied in \cite{Brihaye:2002hr,Brihaye:2002jg,Breitenlohner:2005}, our task
here involved the introduction of a negative cosmological constant. The most
succint way of listing our conclusions is:
\begin{itemize}
\item
The qualitative properties of the regular AAdS solutions in spacetime
dimensions $d=5,6,7,8$ are the same as the corresponding asymptotically
flat ones, namely
\begin{itemize}
\item
A one parameter family of solutions parametrised by the (dimensionless)
gravitational coupling constant $\al$ start at $\al=0$ (the gravity decoupling
limit) and exist up to a maximum $\al_{\rm max}$, after which $\al$ decreases
again and ends at critical value.
\item
For $d=6,7,8$ the value of $\al$ in the second, not that of gravity decoupling,
endpoint becomes very small and stops. In the $d=8$ case it actually vanishes.
%\item
For $d=5$ the value of $\al$ in the second endpoint reaches a finite critical
value, where it does not stop, but oscillates around this critical value.
\item
As long as the physically important $F(2)$ term is present in the YM sector,
there exist no multinode solutions.
\end{itemize}
\item
Black hole counterparts appear to exist for any regular solution.
The qualitative properties of the AAdS black hole solutions are 
similar to the asymptotically flat case:
\begin{itemize}
\item
Different from the four dimensional theory,
the event horizon radius presents a maximal value.
This  maximal value is a function of the gravitational coupling constant $\al$.
\item
For $d=6,7,8$ the black hole solutions form a loop connecting  
the two regular solutions with the same value of $\alpha$.
%\item
The solutions of the five dimensional theory are somehow special, 
presenting a complicated brach structure which depends on $\al$.
\end{itemize}
\end{itemize}

Axially symmetric generalizations of these solutions are likely to exist. We expect them
to share the basic properties of the 
spherically symmetric configurations.
\\
\\
{\bf Acknowledgement}
\\
We are inebted to Dieter Maison for advice and discussions. 
ER  thanks  Cristian Stelea for useful remarks on Section 2.
This work was carried out in the framework of Enterprise--Ireland
Basic Science Research Project SC/2003/390.

%\newpage
%%%%%%%%%%%%%%%%%%%%%%%%%%%%%%%%%%%%%%%%%%%%%%%%%%%%%%%%%%%%%%%%%%%%%%%%%%%%%%

%%%%%%%%%%%%%%%%%%%%%%%%%%%%%%%%%%%%%%%%%%%%%%%%%%%%%%%%%%%%%%%%%%%%%%%%%%%%%%%%%%%%%%%%%

%%%%%%%%%%%%%%%%%%%%%%%%%%%%%%%%%%%%%%%%%%%%%%%%%%%%%%%%%%%%%%%%%%%%%%%%%
\newpage
\textbf{\Large Appendix A: A nonexistence proof for $p=1,~d> 5$ 
finite mass solutions with $\Lambda \leq  0$}
\\
\\
Following the notation used in \cite{Okuyama:2002mh},
we introduce a new variable
\begin{eqnarray}
z=2 \log r,
\end{eqnarray}
and rewrite the basic equations (\ref{eqsF2-1}), (\ref{eqsF2-2}) in the form
\begin{eqnarray}
\label{app1}
&&\frac{dm}{dz}=\frac{\tau_1}{4}e^{\frac{(d-5)}{2}z}\left(4N(\frac{dw}{dz})^2+
\frac{(d-3)}{2}(w^2-1)^2\right),
\\
\label{app2}
&&N \frac{d^2 w}{dz^2}+\left(\frac{(d-5)N}{2}+
(d-3)e^{-\frac{(d-3)}{2}z}\frac{m}{\kappa}+\frac{e^z}{\ell^2}
-\frac{\tau_1e^{-z}}{4\kappa}(d-3)(w^2-1)^2 \right) 
\frac{dw}{dz}=\frac{(d-3)}{4}w(w^2-1),
\end{eqnarray}
where the metric function $\sigma$ has been eliminated by using
(\ref{eqsF2-3}). The function $N$ is given by (\ref{N}), namely as
 $N=1-\frac{2}{\kappa}e^{-(d-3)z/2}m(z)+ e^z/\ell^2$.

To devise a proof for the nonexistence 
of finite mass solutions of the above system, it is convenient
to introduce the function
\begin{eqnarray}
\label{defE}
E=\frac{1}{2}N\left(\frac{dw}{dz}\right)^2-\frac{(d-3)}{16}(w^2-1)^2,
\end{eqnarray}
which, from (\ref{app1}) and (\ref{app2}) satisfies the equation
\begin{eqnarray}
\label{eqE}
\frac{dE}{dz}= -\left(\frac{dw}{dz}\right)^2\left(\frac{\tau_1}{\kappa}
e^{-z}E+\frac{e^z}{\ell^2}+
\frac{m}{\kappa}(d-3)e^{-\frac{(d-3)}{2}z}\right).
\end{eqnarray}
 The relations in Appendix B of Ref.
\cite{Okuyama:2002mh} are recovered for $d=5$, where a nonexistence proof
is presented for particle-like solutions with finite mass (the extension
to black hole case is considered in \cite{Brihaye:2005tx}).
Therefore in what follows we will take $d>5$ only.
 
The approximate form of the function $E$ at the origin (or event
horizon) and infinity is found by taking $P=1$
in relations (\ref{r=0}), (\ref{eh}) and (\ref{29}) 
given in Section 3 and replacing in (\ref{defE}).
It is obvious that $E(r_h)<0$, since $N(r_h)=0$; at the same time, the corresponding expression
as $r \to 0$ ($i.e.$ $z \to -\infty$) is
\begin{eqnarray}
\label{E-reg}
E\simeq\frac{1}{4}(5-d)b^2e^{2z}+\dots,
\end{eqnarray}
and we find $E\to -0$ in this limit.

Also, the relations
(\ref{r=0}), (\ref{eh}) give $m(r=0)>0$, $m(r_h)>0$, 
which, together with the equation (\ref{app1})
implies that the mass function $m(r)$ is positive definite.

Besides, by replacing in (\ref{defE}) the asymptotic expressions (\ref{29}) as
$r \to \infty$ it follows that 
\begin{eqnarray}
\label{E-inf}
E\simeq\frac{w_1^2(d-3)^2}{8\ell^2}e^{-\frac{(d-4)}{2}z}+\frac{1}{8}w_1^2(d-3)(d-5)e^{-(d-3)z}+\dots,
\end{eqnarray}
$i.e.$ $E\to +0$ as $z \to \infty$, for finite mass solutions.
 
Therefore, if the solution is regular everywhere, $E$ must vanish at some
finite point $z_0$, and $dE/dz \geq 0$ there, with $E>0$ for $z>z_0$ 
(when there are several positive roots of $E$,
we take the largest one).
However, another point should exist $z_1>z_0$ such that  $dE/dz = 0$ $i.e.$
the function
$E$ should present a positive maximum for some value of $z$.
Now we integrate the equation (\ref{eqE}) between $z_0$ and $z_1$ and find
\begin{eqnarray}
E(z_1)=-\int_{r_0}^{r_1}\left(\frac{dw}{dz}\right)^2\left(\frac{\tau_1}{\kappa}
e^{-z}E+\frac{e^z}{\ell^2}+
\frac{m}{\kappa}(d-3)e^{-\frac{(d-3)}{2}z}\right)dz<0,
\end{eqnarray}
which contradicts $E(z_1)>0$. Therefore $E(z)$ should vanish identically and 
one finds no $d> 5$ finite mass, spherically symmetric EYM configurations
in a $F(2)$ theory.
Note that this argument does not exclude the existence of configuration with
a diverging mass functions as $r \to \infty$.
\\
\\
\newpage
%%%%%%%%%%%%%%%%%%%%%%%%%%%%%%%%%%%%%%%%%%%%%%%%%%%%%%%%%%%%%%%%%%%%%%%%%%%%%%
\textbf{\Large Appendix B: A matter counterterm proposal}
%%%%%%%%%%%%%%%%%%%%%%%%%%%%%%%%%%%%%%%%%%%%%%%%%%%%%%%%%%%%%%%%%%%%%%%%%%%%%%
\\
\\
In this Section we comment on the issue of mass definition
of  AAdS solutions in a $F(2)^2$ (i.e. $p=1$) theory, if we do not exercise
the option of employing higher order YM survature terms.
As proven in Section 2, although the spacetime is still AAdS,
the mass function $m(r)$ of these solutions generically diverges as
$r^{d-5}$ (or as $\log r$ for $d=5$).
AAdS solutions with a diverging ADM mass have been considered recently by some
authors, mainly for a scalar field in the bulk
(see e.g. \cite{Hertog:2004dr}-\cite{Liu:2004it}. In this case it might be
possible to relax the standard asymptotic
conditions without loosing the original symmetries,
but modifying the charges in order to take into account the presence of matter
fields.

Similar to the case of  scalar field, for $d>5$
it is still  possible to obtain a finite mass 
of EYM solutions in a $F^2$ theory by
allowing $I_{ct}$ to depend not only on
the boundary metric $h _{AB}$, but
also on the gauge field 
strength tensor.
This means that the
quasilocal stress-energy tensor (\ref{TAB}) also
acquires a contribution coming from the matter fields.

We find that by adding a  counterterm of the form
\begin{eqnarray}
\label{Ict-mat}
I_{ct}^{(m) }=
-\frac{1}{(d-5)}
 \int_{\partial M}d^{d-1}x\sqrt{-h }
 ~{\rm tr}~ F_{AB}F^{AB}  
\end{eqnarray}
to the expression (\ref{ct}),
the divergence disappears.
This yields a supplementary contribution to (\ref{TAB})
\begin{eqnarray}
\label{TAB-mat}
T_{AB}=-\frac{1}{8\pi G}\frac{1}{d-5}h_{AB}~{\rm tr}~F_{CD}F^{CD}. 
\end{eqnarray}
The mass of the $d>5$ solutions computed in this way 
is
\begin{eqnarray}
\label{mass-ctm}
E=\frac{(d-2)\Omega_{d-2}}{8 \pi G} M_0+E_0 
\end{eqnarray}
where $M_0$ is the constant appearing in the asymptotic
expansion (\ref{mass-div}).
It can also be proven that this prescription
leads to a finite action
and the entropy-area relation is satisfied.
However, as seen in Figure 2, the parameter $M_0$ of the $p=1$ solutions takes
negative values, pointing to some pathological properties. 

It would be nice to have a rigorous derivation of the matter counterterm
expression, possibly along the lines of Ref. \cite{Batrachenko:2004fd}.
Also, there remains the issue of the $d=5$ solutions, whose
logarithmic divergeces require a different approach.

\newpage
\setlength{\unitlength}{1cm}

\begin{picture}(16,16)
\centering
\put(-1,0){\epsfig{file=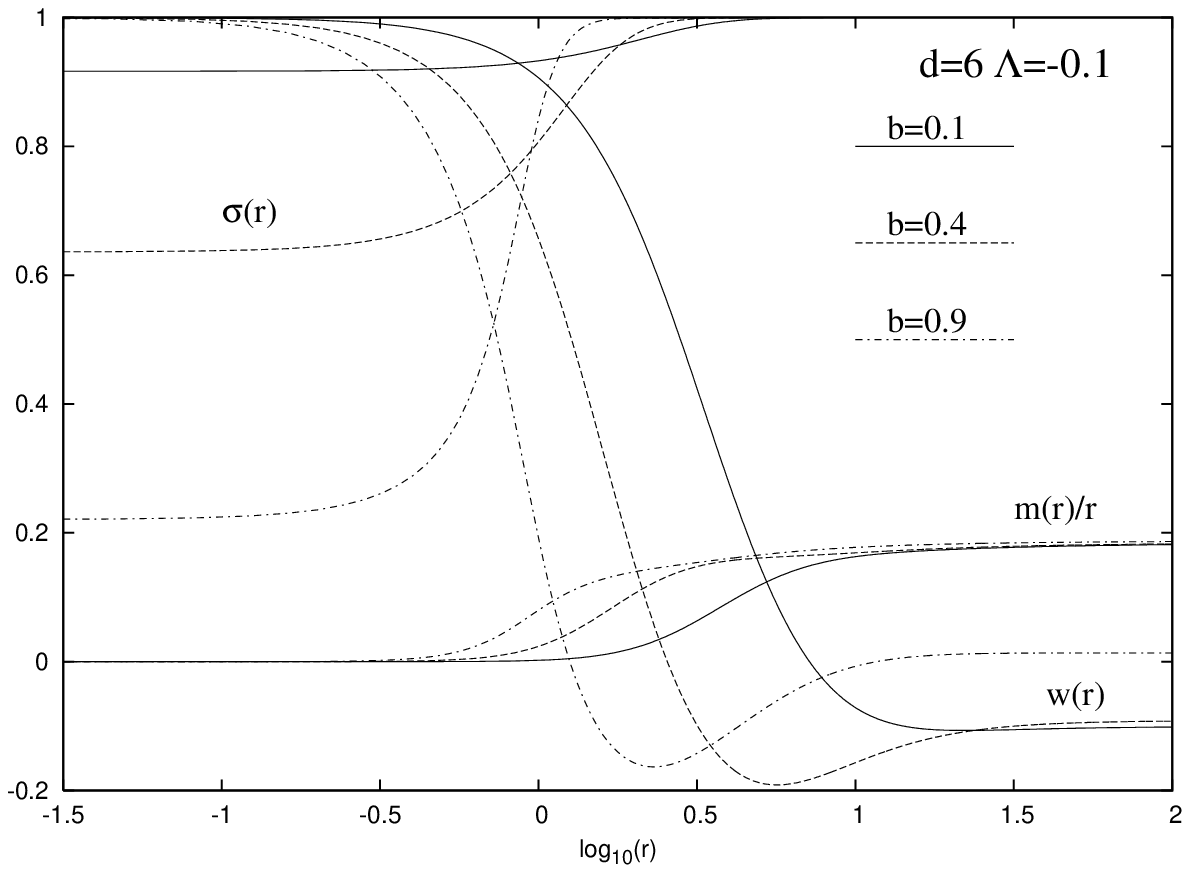,width=16cm}}
%\caption{a}
\end{picture}
\\
\\
\\
{\small {\bf Figure 1.}} 
The functions $\sigma(r)$, $w(r)$ and the ratio $m(r)/r$
are plotted as functions of radius
for typical $d=6$ regular solutions in a $F^2$ EYM theory with $\Lambda=-0.1$ 
and several values of the parameter $b=-\frac12w''(0)$.
%%%%%%%%%%%%%%%%%%%%%%%%%
\newpage
\setlength{\unitlength}{1cm}

\begin{picture}(16,16)
\centering
\put(-1,0){\epsfig{file=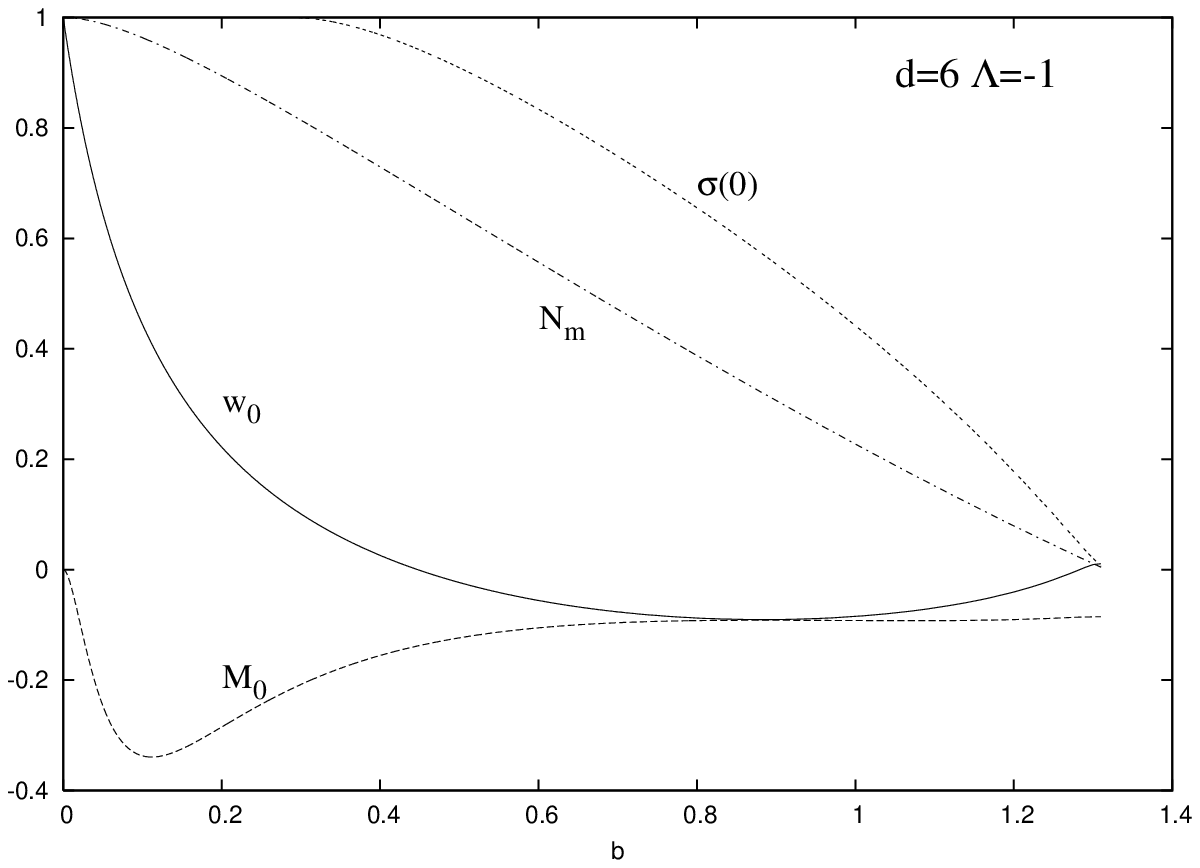,width=18cm}}
%\caption{a}
\end{picture}
\\
\\
\\
{\small {\bf Figure 2.}} 
The value $N_m$ of the minimum of the metric function $N(r)$, 
the  parameter $M_0$ appearing in the asymptotic expansion of $m(r)$,
the asymptotic value $w_0$ of the gauge function $w(r)$ as
well as the value $\sigma(0)$ of the metric function $\sigma$ at the origin,
are shown as functions of $b$ for $d=6$ solutions of $F^2$ theory with $\Lambda=-1$.
%%%%%%%%%%%%%%%%%%%%%%%%%
\newpage
\setlength{\unitlength}{1cm}

\begin{picture}(16,16)
\centering
\put(-1,0){\epsfig{file=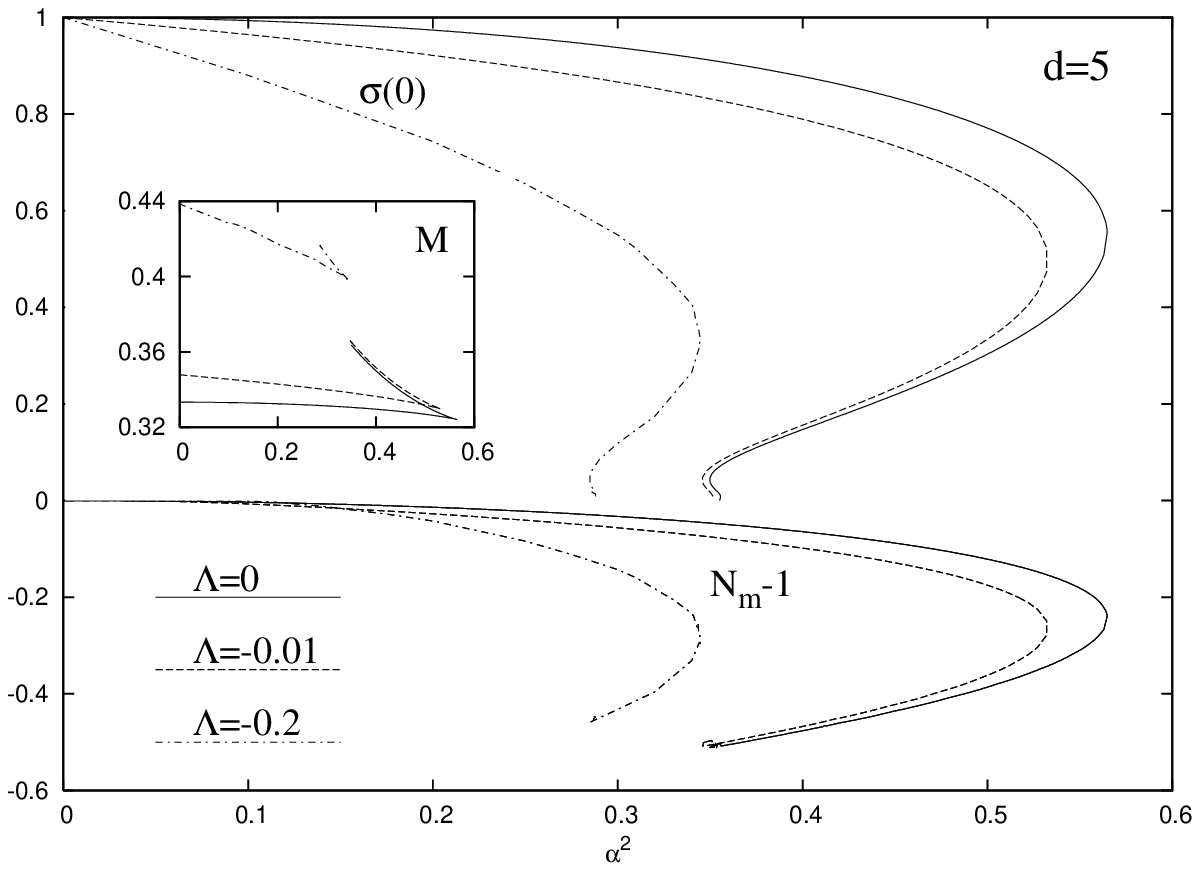,width=18cm}}
%\caption{a}
\end{picture}
\\
\\
\\
{\small {\bf Figure 3.}} 
The value $N_m$ of the minimum of the metric function $N(r)$,
the mass parameter $M$ as
well as the value of the metric function $\sigma$ at the origin, $\sigma(0)$,
are shown for $d=5$ solutions
as functions of $\alpha^2$ and several values of $\Lambda$.

%%%%%%%%%%%%%%%%%%%%%%%%%

\newpage
\setlength{\unitlength}{1cm}

\begin{picture}(16,16)
\centering
\put(-1,0){\epsfig{file=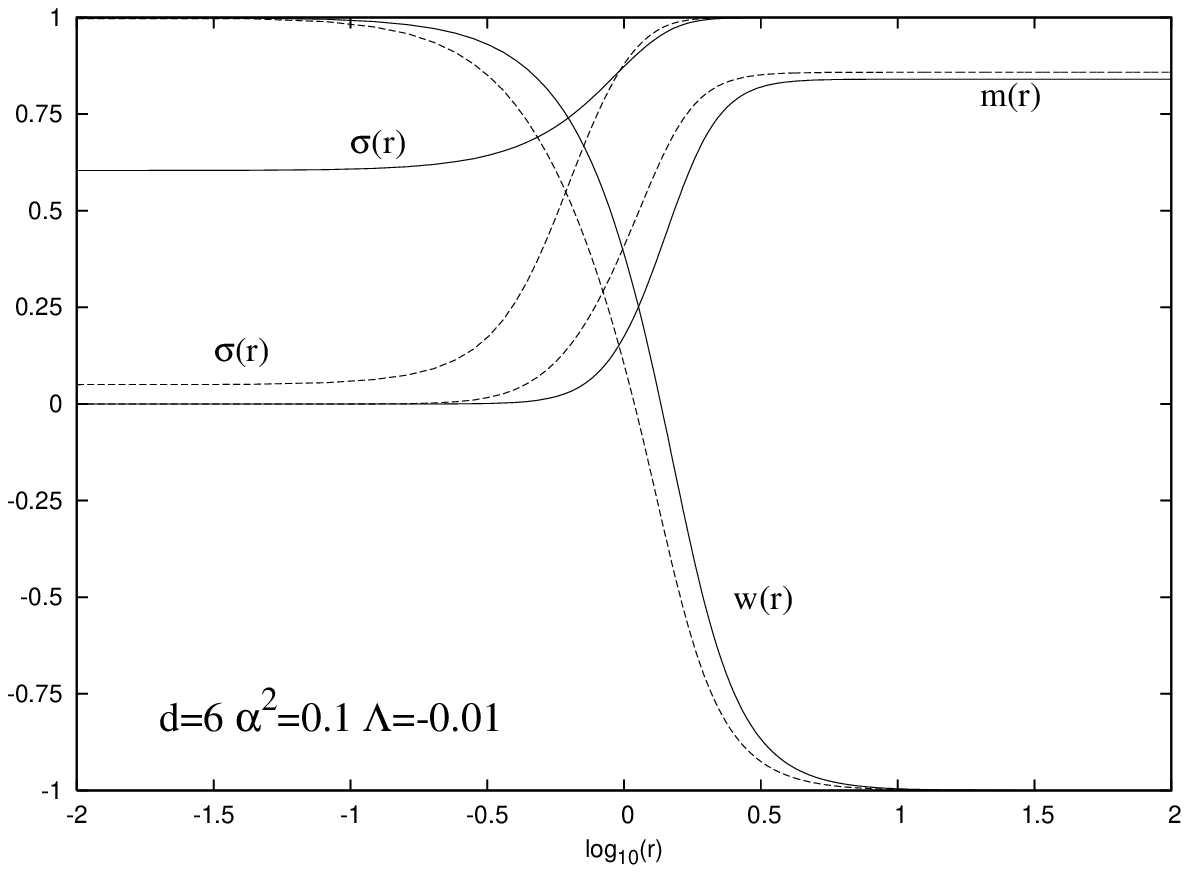,width=16cm}}
%\caption{a}
\end{picture}
\\
\\
\\
{\small {\bf Figure 4.}} 
The profiles of the functions $m(r),~\sigma(r)$ and $w(r)$  
are plotted as functions of radius
for typical $d=6$ regular solutions in a  
EYM theory with $p=1,2$ terms and $\alpha^2=0.1,~\Lambda=-0.01$.
Here and in Figure 14 the continuos line corresponds to an upper branch solution,
the dotted line denoting lower branch  profiles.
%%%%%%%%%%%%%%%%%%%%%%%%%
\newpage
\setlength{\unitlength}{1cm}

\begin{picture}(16,16)
\centering
\put(-1,0){\epsfig{file=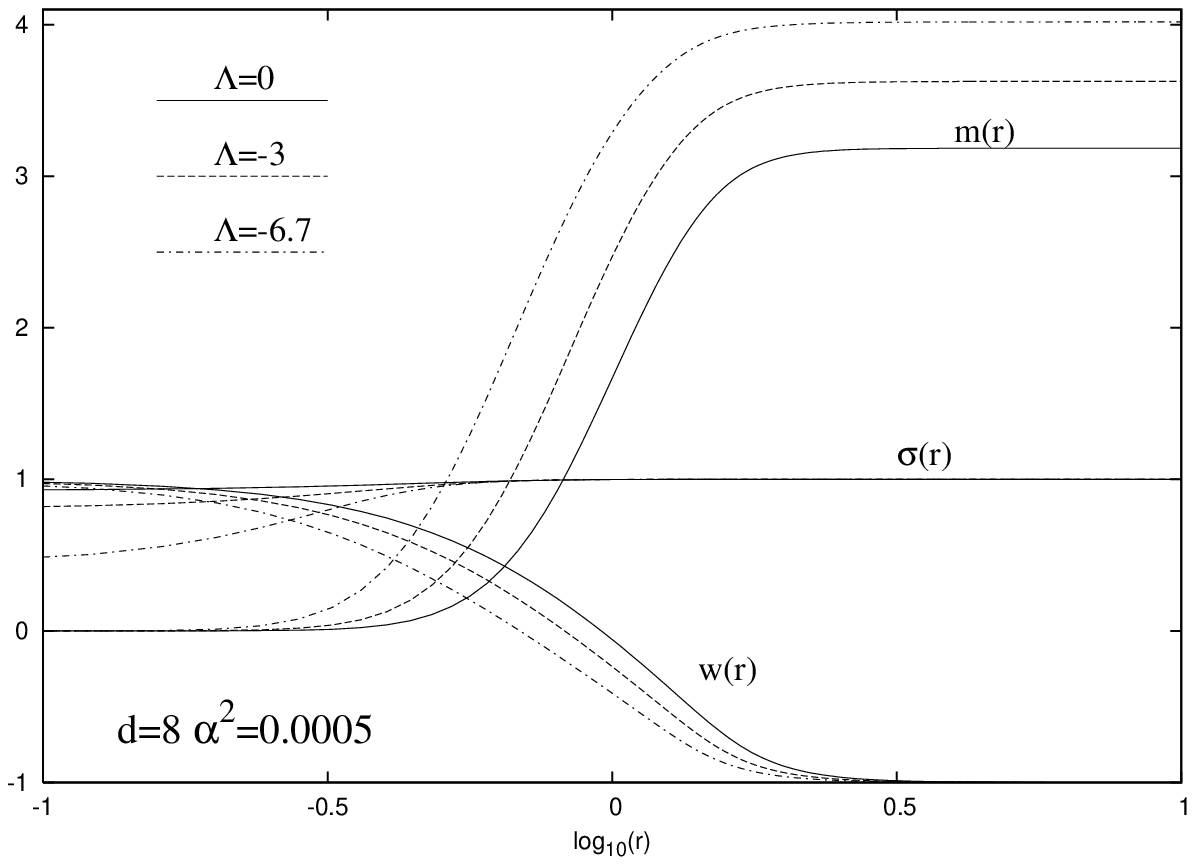,width=16cm}}
%\caption{a}
\end{picture}
\\
\\
\\
{\small {\bf Figure 5.}} 
Typical globally regular solutions of $d=8$ EYM theory with $\alpha^2=0.0005$ and $p=1,2$ terms
are plotted as functions of radius for several values of $\Lambda$. 

%%%%%%%%%%%%%%%%%%%%%%%%%
\newpage
\setlength{\unitlength}{1cm}

\begin{picture}(16,16)
\centering
\put(-1,0){\epsfig{file=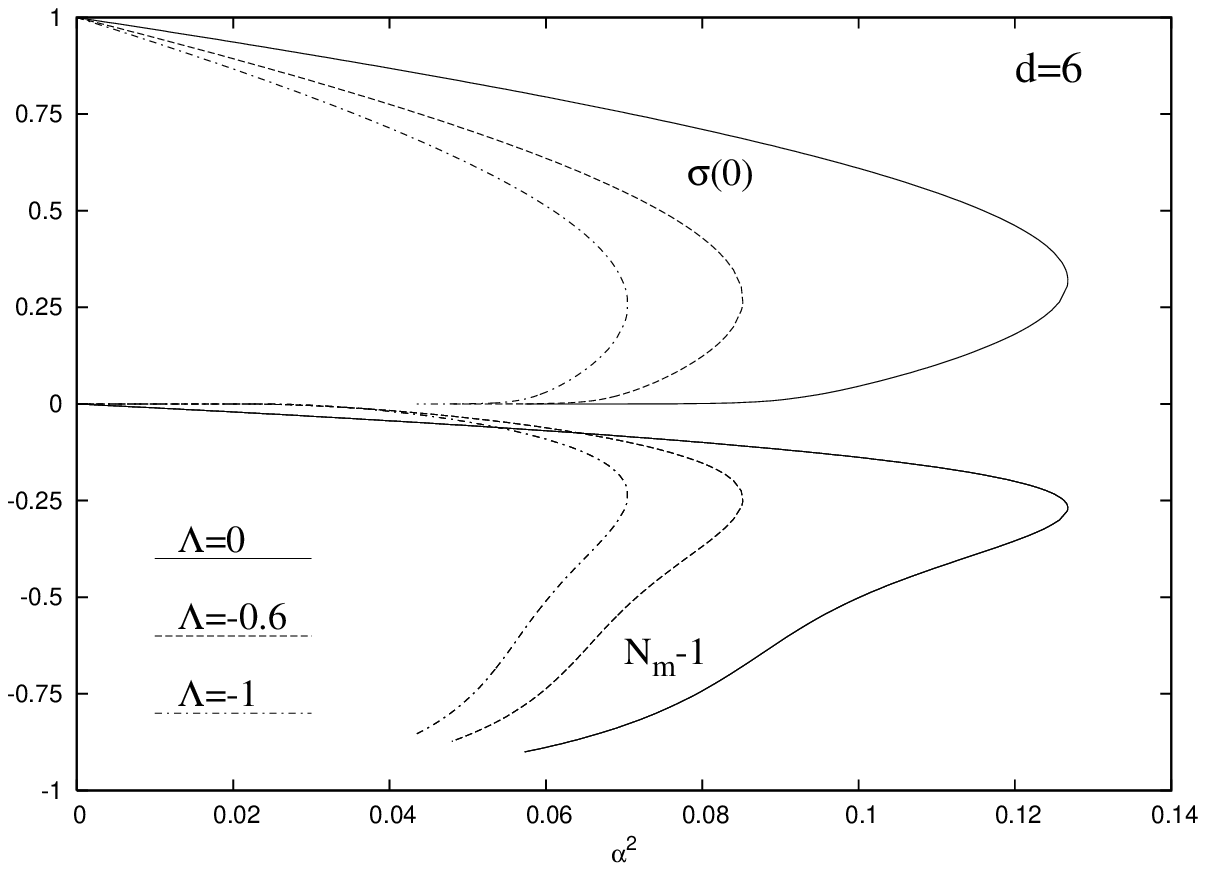,width=16cm}}
%\caption{a}
\end{picture} 
\\
\\
\\
\begin{center}
Figure 6a.
\end{center}
%%%%%%%%%%%%%%%%%%%%%%%%%

\newpage
\setlength{\unitlength}{1cm}

\begin{picture}(16,16)
\centering
\put(-2,0){\epsfig{file=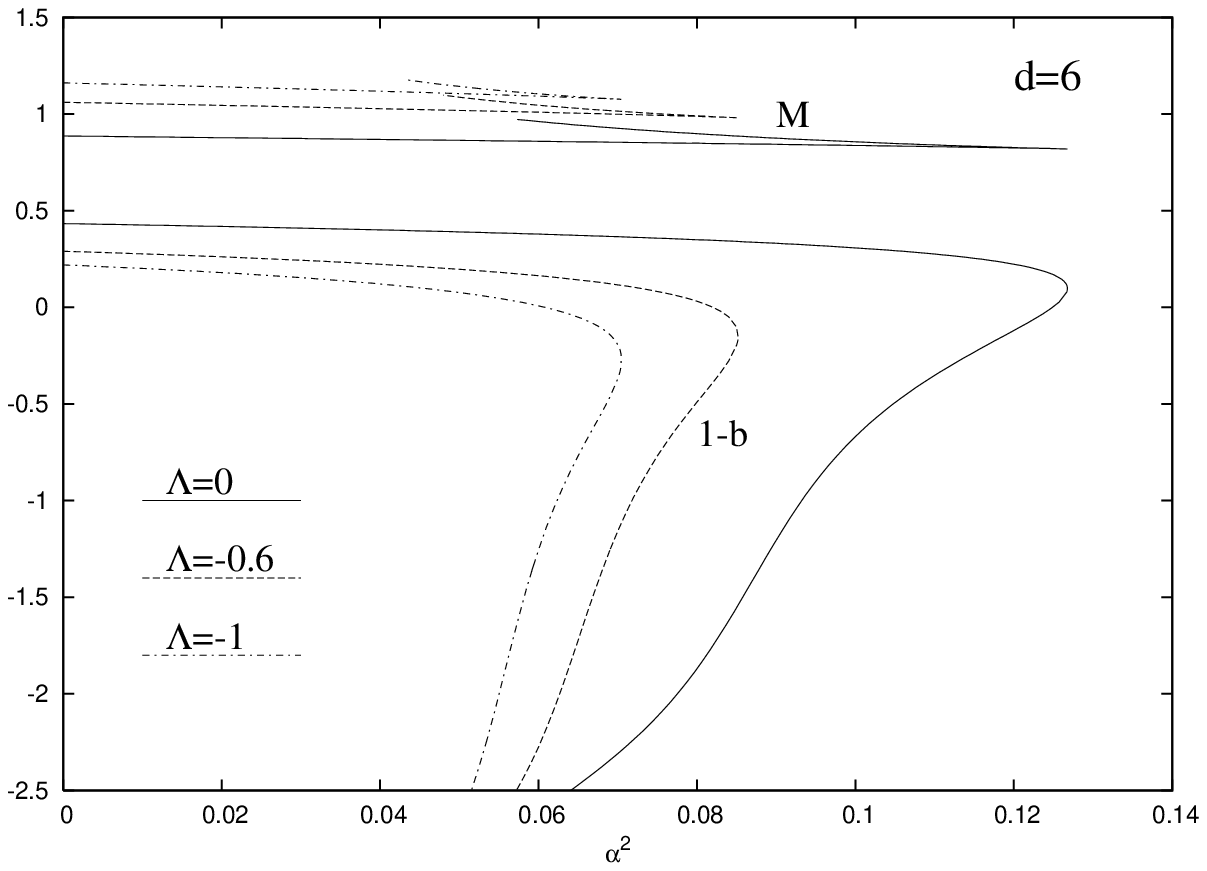,width=16cm}}
%\caption{a}
\end{picture}
\\
\\
\\
\begin{center}
Figure 6b.
\end{center}
{\small {\bf Figure 6}} 
The value $N_m$ of the minimum of the metric function $N(r)$ and 
the value of the metric function $\sigma$ at the origin $\sigma(0)$ (Figure 6a), 
and, the parameters  $M$ and $b$ (Figure 6b),
are shown for $d=6$ solutions
as functions of $\alpha^2$ and several values of $\Lambda$.
%%%%%%%%%%%%%%%%%%%%%%%%%

\newpage
\setlength{\unitlength}{1cm}

\begin{picture}(16,16)
\centering
\put(-1,0){\epsfig{file=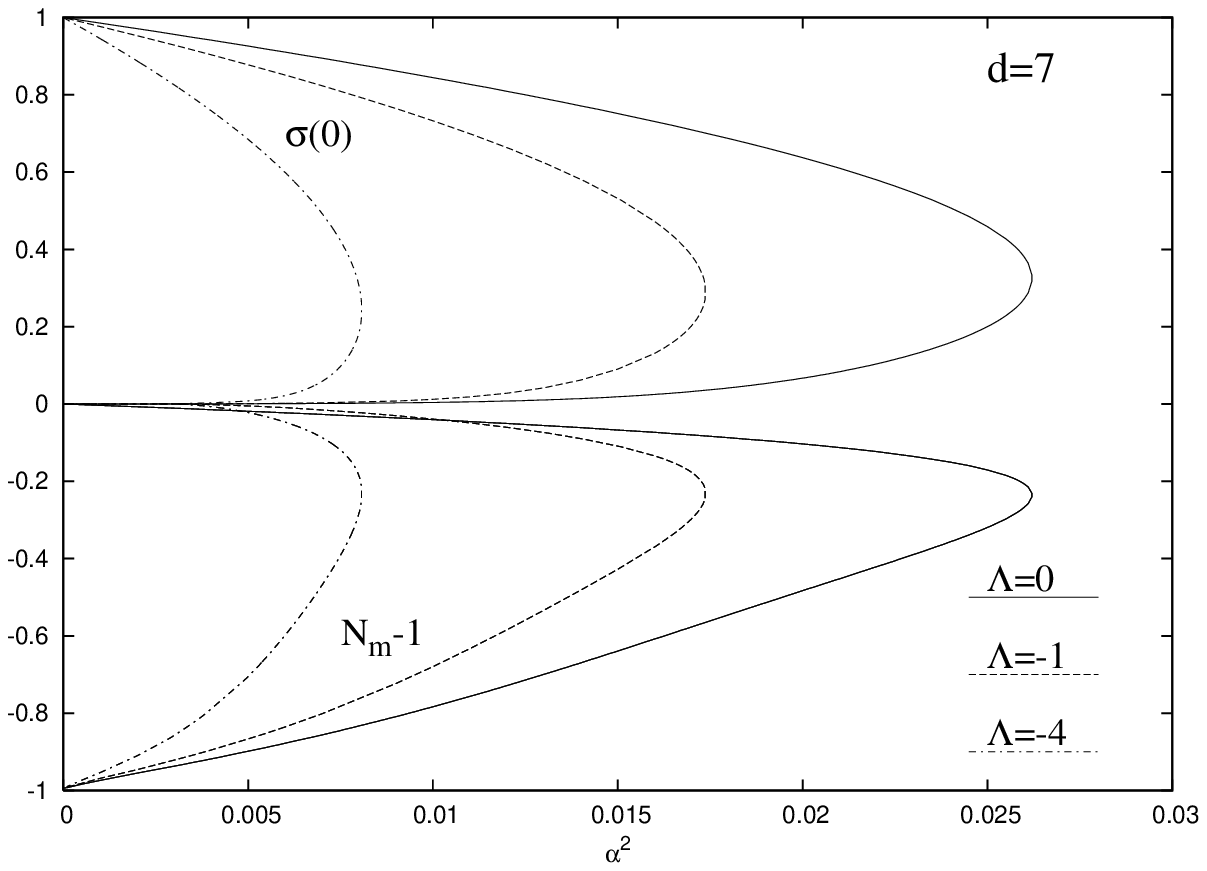,width=16cm}}
%\caption{a}
\end{picture} 
\\
\\
\\
\begin{center}
Figure 7a.
\end{center}
%%%%%%%%%%%%%%%%%%%%%%%%%

\newpage
\setlength{\unitlength}{1cm}

\begin{picture}(16,16)
\centering
\put(-1,0){\epsfig{file=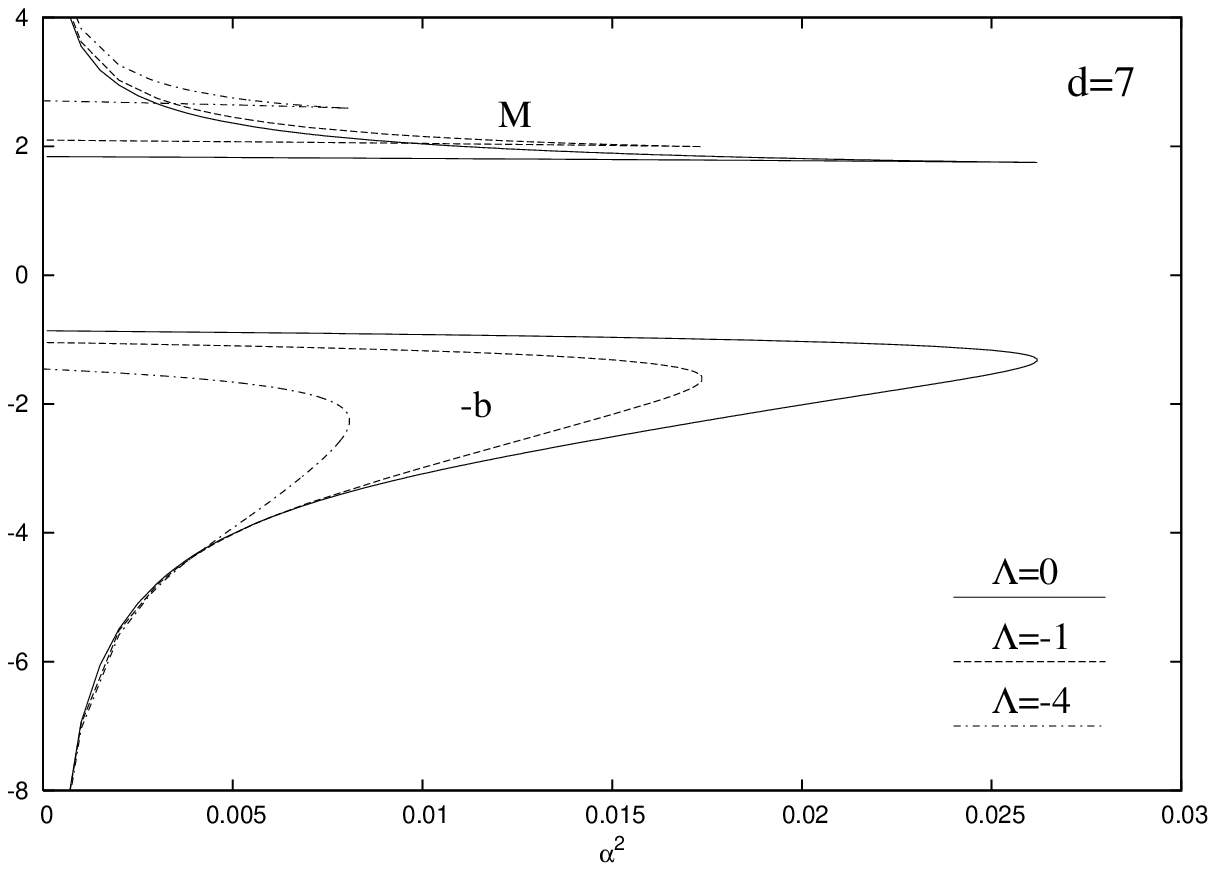,width=16cm}}
%\caption{a}
\end{picture}
\\
\\
\\
\begin{center}
Figure 7b.
\end{center}
{\small {\bf Figure 7}} 
The value $N_m$ of the minimum of the metric function $N(r)$ and 
the value of the metric function $\sigma$ at the origin $\sigma(0)$ (Figure 7a), 
and, the parameters $M$ and $b$ (Figure 7b),
are shown for $d=7$ solutions
as functions of $\alpha^2$ and several values of $\Lambda$.
%%%%%%%%%%%%%%%%%%%%%%%%%
\newpage
\setlength{\unitlength}{1cm}

\begin{picture}(16,16)
\centering
\put(-1,0){\epsfig{file=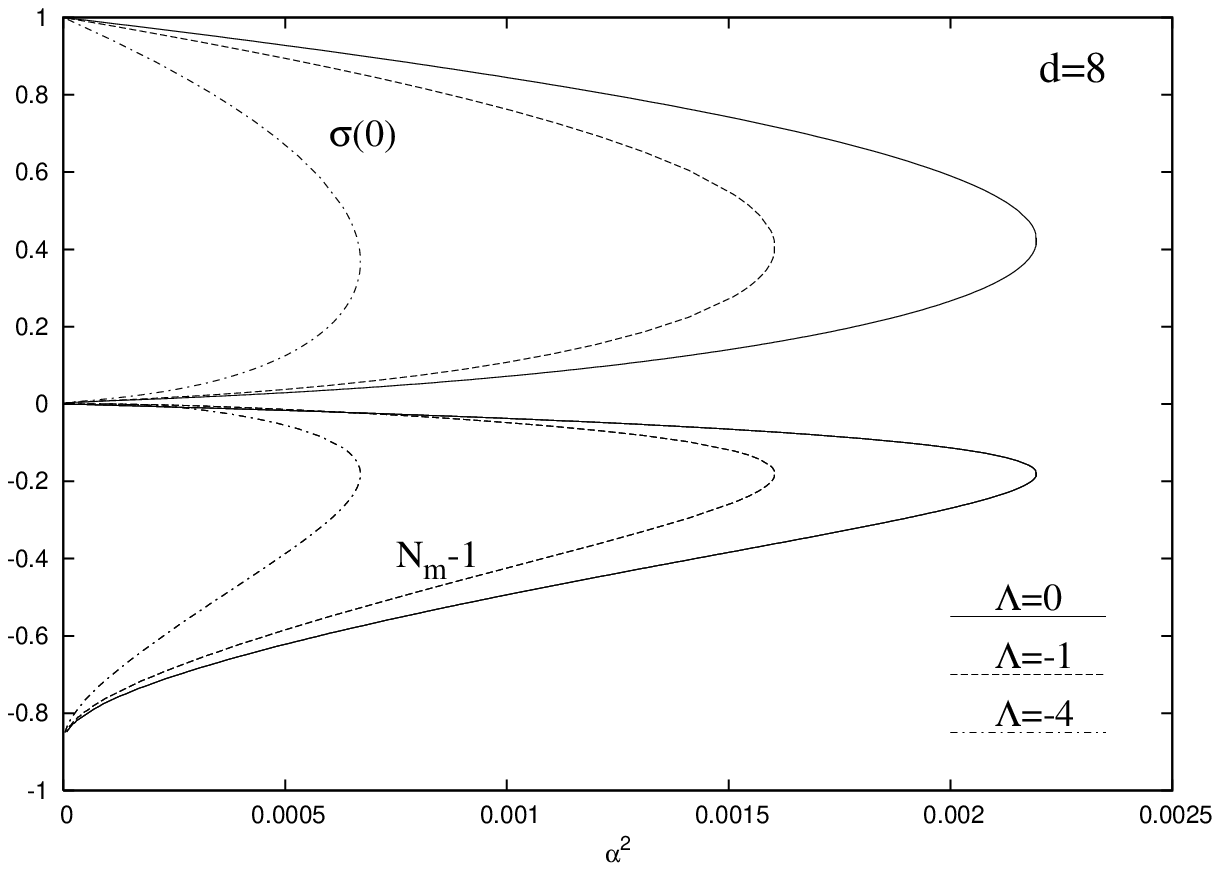,width=16cm}}
%\caption{a}
\end{picture} 
\\
\\
\\
\begin{center}
Figure 8a.
\end{center}
%%%%%%%%%%%%%%%%%%%%%%%%%
\newpage
\setlength{\unitlength}{1cm}

\begin{picture}(16,16)
\centering
\put(-1,0){\epsfig{file=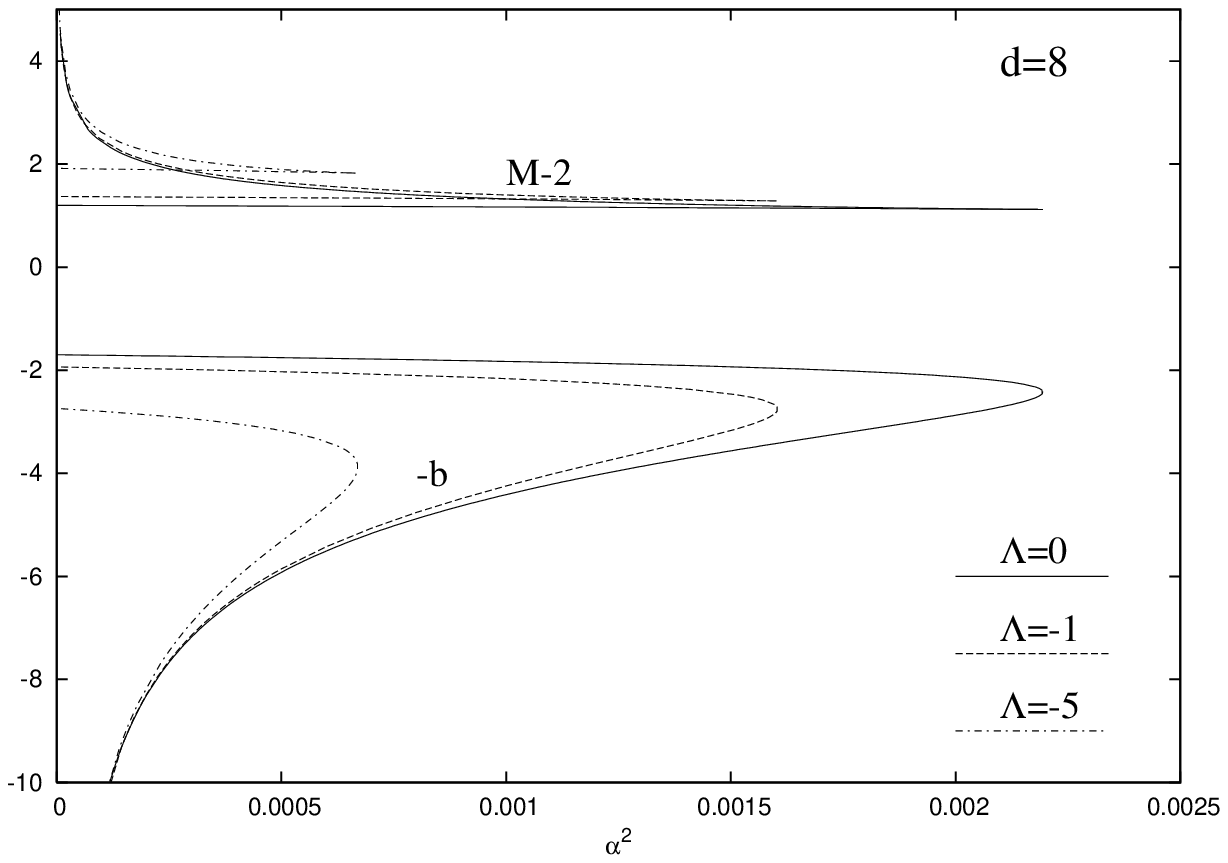,width=16cm}}
%\caption{a}
\end{picture}
\\
\\
\\
\begin{center}
Figure 8b.
\end{center}
{\small {\bf Figure 8}} 
The value $N_m$ of the minimum of the metric function $N(r)$ and 
the value of the metric function $\sigma$ at the origin $\sigma(0)$ (Figure 8a), 
and the parameters $M$ and $b$ (Figure 8b),
are shown for $d=8$ solutions
as functions of $\alpha^2$ and several values of $\Lambda$.
%%%%%%%%%%%%%%%%%%%%%%%%%

\newpage
\setlength{\unitlength}{1cm}

\begin{picture}(16,16)
\centering
\put(-1,0){\epsfig{file=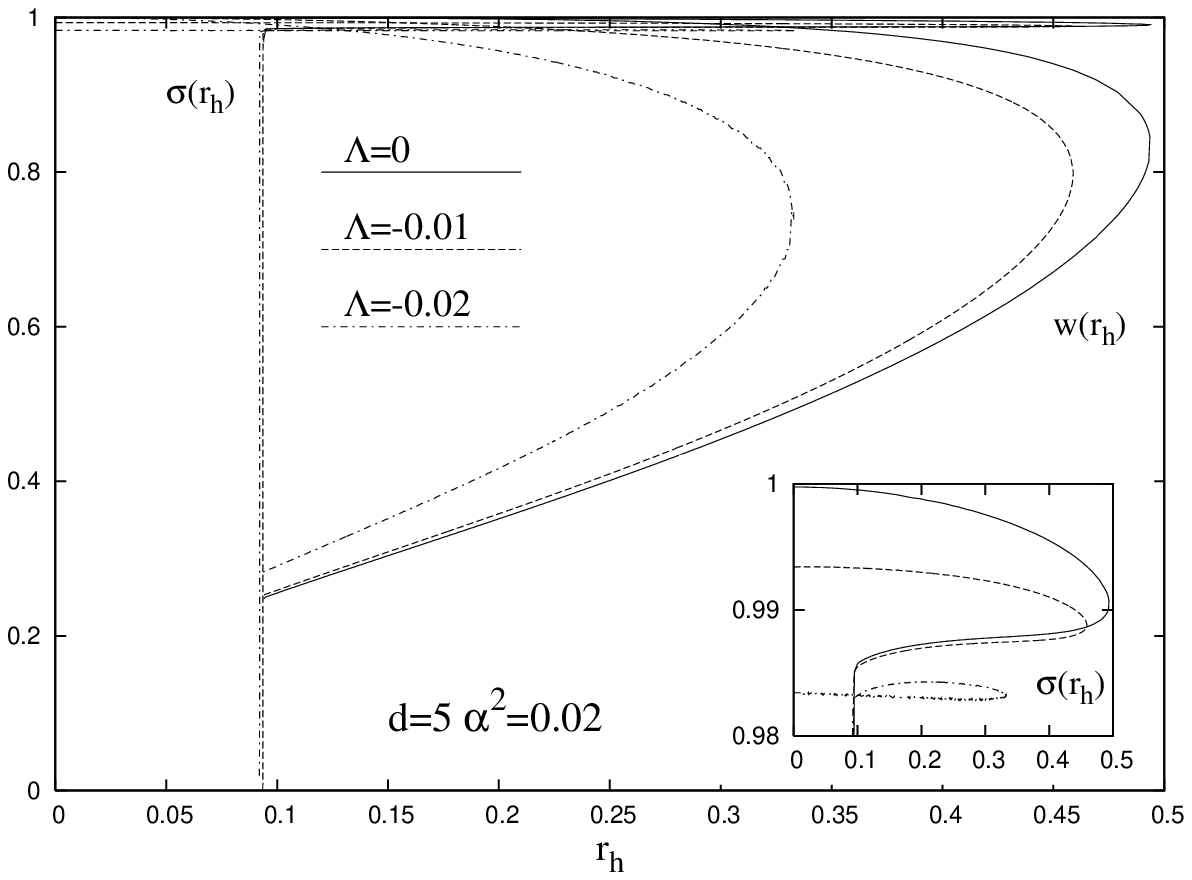,width=16cm}}
%\caption{a}
\end{picture}
\\
\\
\\
\\
\\
\\
\begin{center}
Figure 9a.
\end{center}
%%%%%%%%%%%%%%%%%%%%%%%%%
\newpage
\setlength{\unitlength}{1cm}

\begin{picture}(16,16)
\centering
\put(-1,0){\epsfig{file=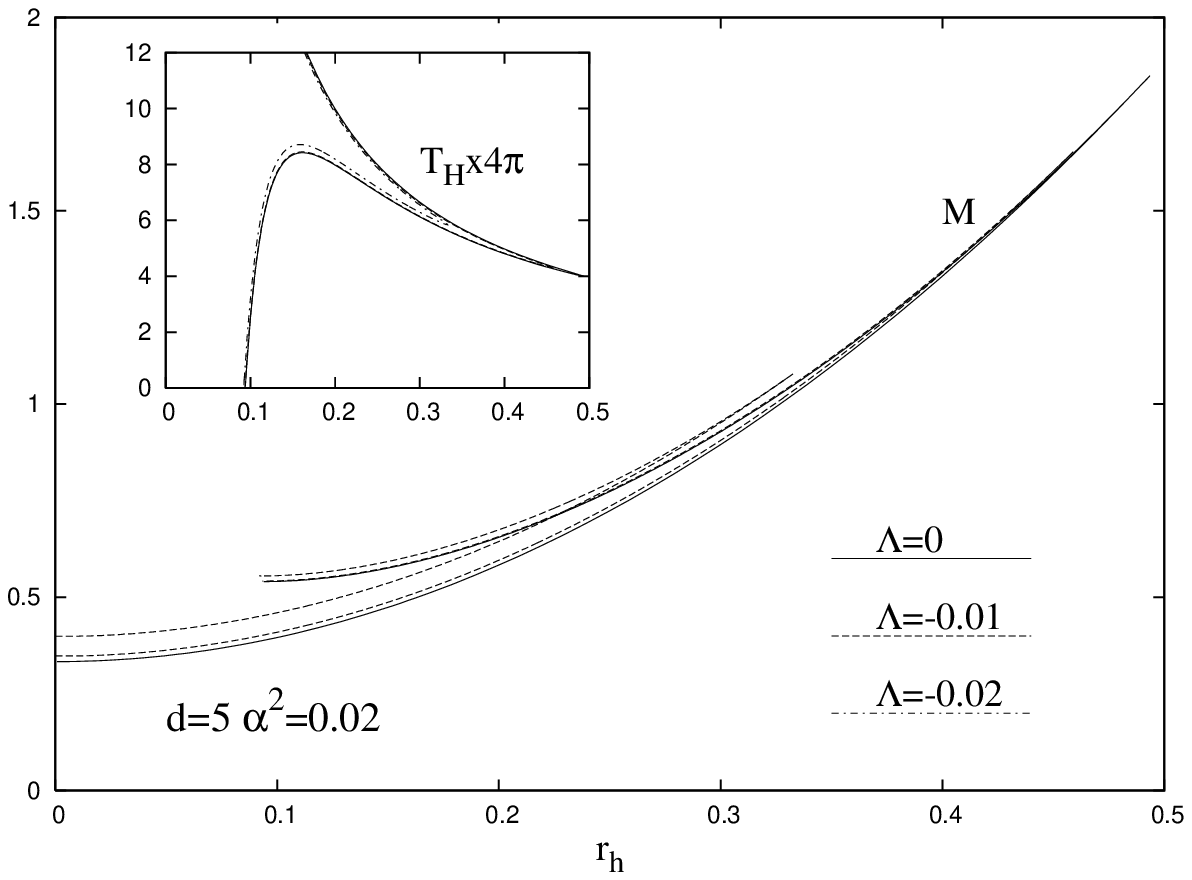,width=16cm}}
%\caption{a}
\end{picture}
\\
\\
\\
\\
\\
\\
\begin{center}
Figure 9b.
\end{center}
{\small {\bf Figure 9}} 
In Figure 9a we plot the value of the gauge field 
function at the horizon $w(r_h)$ and
$\sigma(r_h)$, the value of the metric function $\sigma$ at the horizon 
(the magnified profiles of $\sigma(r_h)$ are displayed in the
window, to help distinguish these from the profiles of $w(r_h)$).
In Figure 9b, the mass parameter $M$
and the Hawking temperature $T_H$ are presented. All profiles
as functions of the event horizon radius $r_h$ for
the $p=1,~2$ black hole solutions in five dimensions with 
$\alpha^2=0.02$ and several values of $\Lambda$. 
%%%%%%%%%%%%%%%%%%%%%%%%%

%%%%%%%%%%%%%%%%%%%%%%%%%
\begin{picture}(16,16)
\centering
\put(-1,0){\epsfig{file=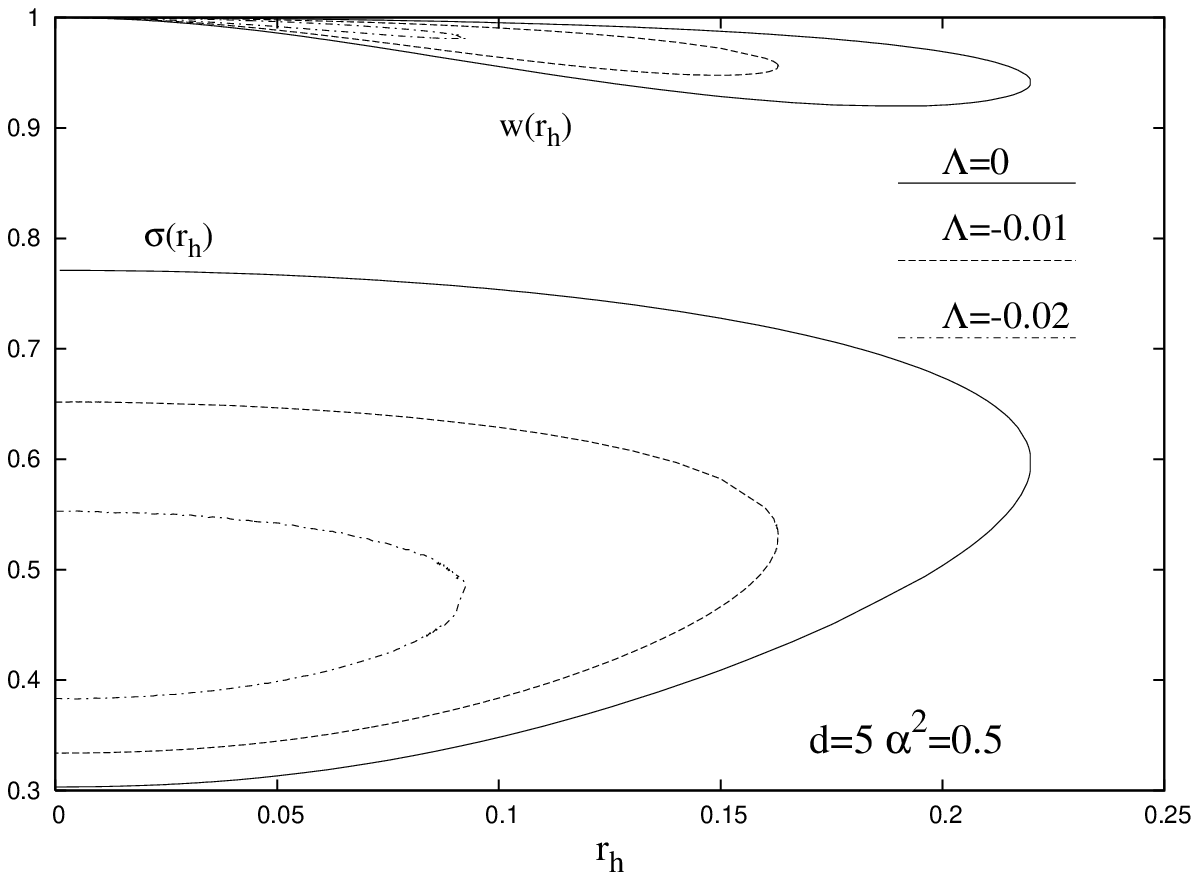,width=16cm}}
%\caption{a}
\end{picture}
\\
\\
\\
\\
\\
\\
\begin{center}
Figure 10a.
\end{center}
%%%%%%%%%%%%%%%%%%%%%%%%%
\newpage
\setlength{\unitlength}{1cm}

\begin{picture}(16,16)
\centering
\put(-1,0){\epsfig{file=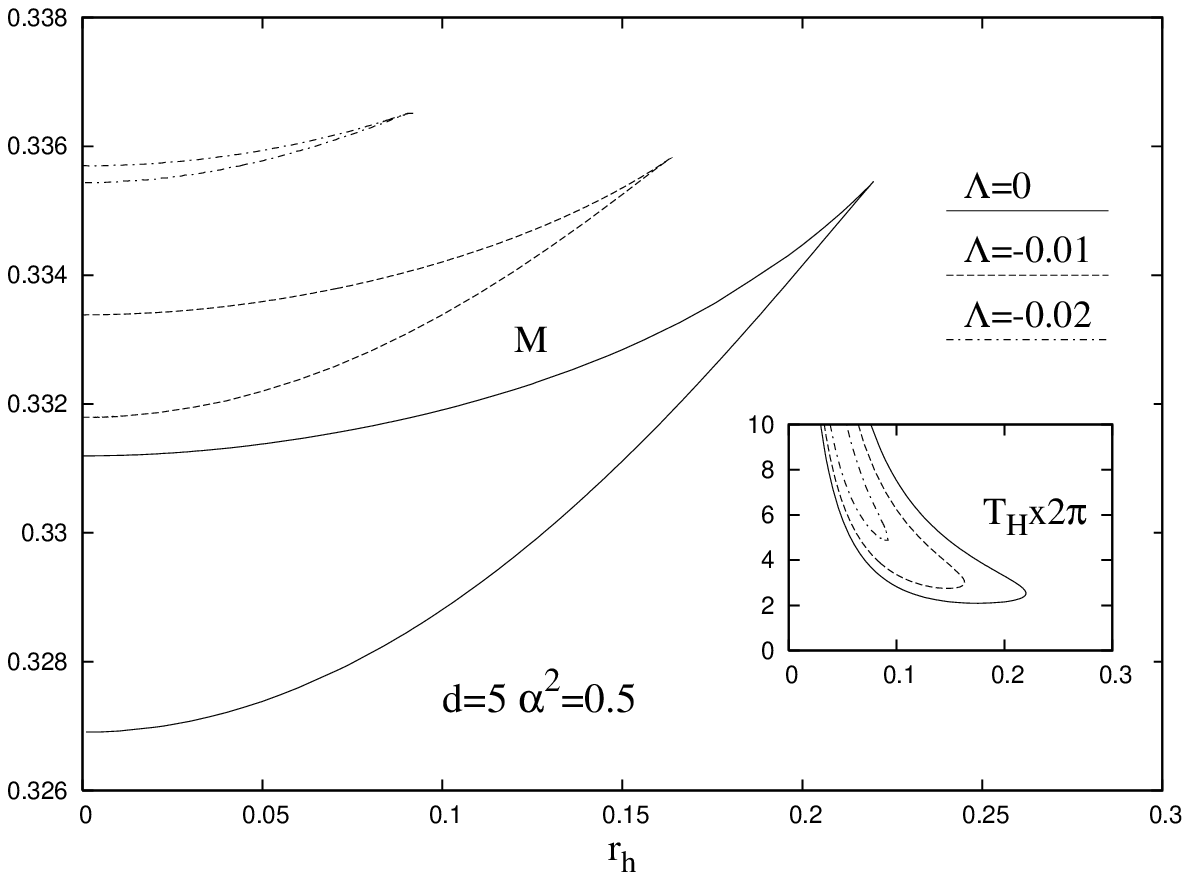,width=16cm}}
%\caption{a}
\end{picture}
\\
\\
\\
\\
\\
\\
\begin{center}
Figure 10b.
\end{center}
{\small {\bf Figure 10}} 
The value of the gauge field function at the horizon $w(r_h)$ and
$\sigma(r_h)$, the value of the metric function $\sigma$ at the horizon, 
(Figure 10a), as well as the mass parameter $M$
and the Hawking temperature $T_H$  (Figure 10b) 
are shown as functions of the event horizon radius $r_h$ for
the $p=1,~2$ black hole solutions in five dimensions with 
$\alpha^2=0.5$ and several values of $\Lambda$.

%%%%%%%%%%%%%%%%%%%%%%%%%
\newpage
\setlength{\unitlength}{1cm}

\begin{picture}(16,16)
\centering
\put(-1,0){\epsfig{file=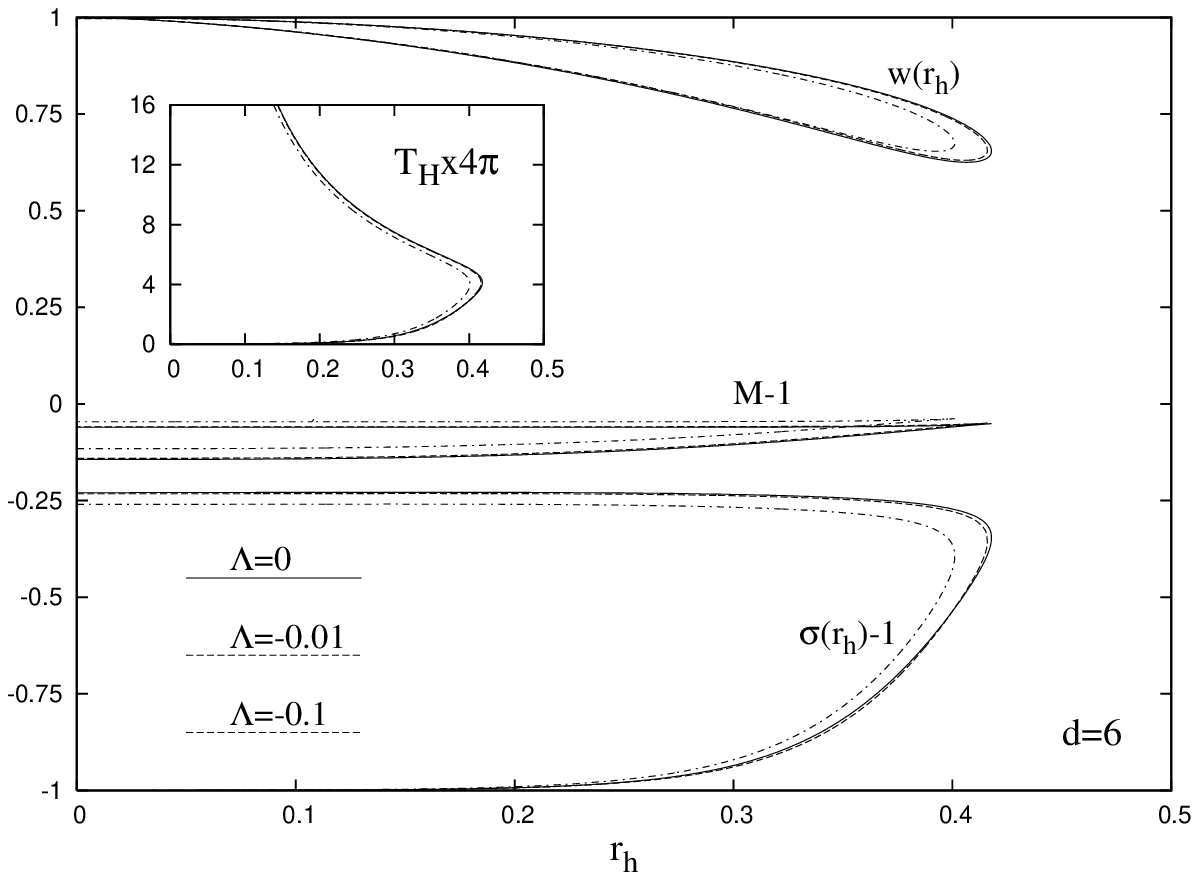,width=16cm}}
%\caption{a}
\end{picture}
\\
\\
\\
{\small {\bf Figure 11.}} 
The value of the gauge field function at the horizon $w(r_h)$ and
$\sigma(r_h)$, the value of the metric function $\sigma$ at the horizon,  
as well as the mass parameter $M$
and the Hawking temperature $T_H$  
are shown as functions of the event horizon radius $r_h$ for
the $p=1,~2$  black hole solutions in six dimensions with 
$\alpha^2=0.066$ and several values of $\Lambda$. 
%%%%%%%%%%%%%%%%%%%%%%%%%
\newpage
\setlength{\unitlength}{1cm}

\begin{picture}(16,16)
\centering
\put(-1,0){\epsfig{file=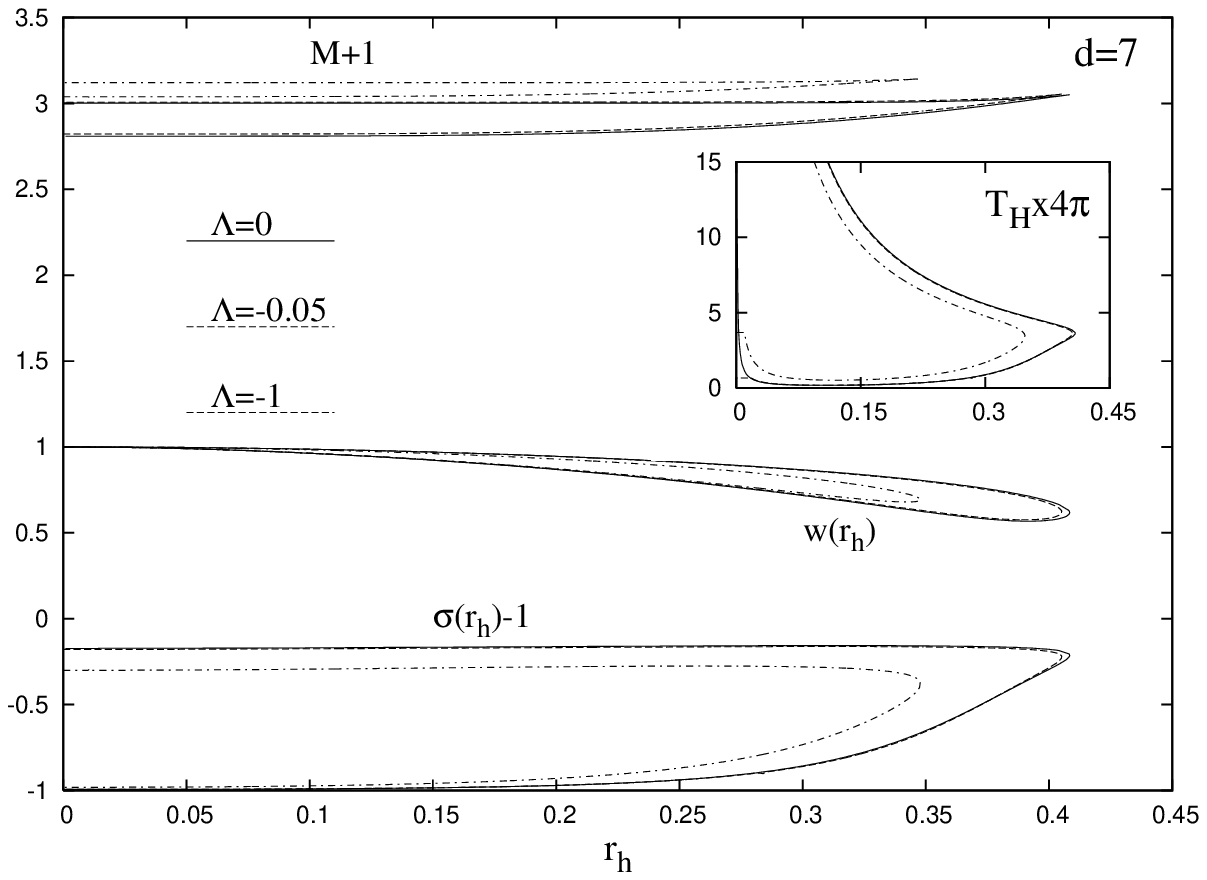,width=16cm}}
%\caption{a}
\end{picture}
\\
\\
\\
{\small {\bf Figure 12.}} 
The value of the gauge field function at the horizon $w(r_h)$ and
$\sigma(r_h)$, the value of the metric function $\sigma$ at the horizon, 
as well as the mass parameter $M$
and the Hawking temperature $T_H$  
are shown as functions of the event horizon radius $r_h$ for
the $p=1,~2$  black hole solutions in seven dimensions with 
$\alpha^2=0.011$ and several values of $\Lambda$. 
%%%%%%%%%%%%%%%%%%%%%%%%%
\newpage
\setlength{\unitlength}{1cm}

\begin{picture}(16,16)
\centering
\put(-1,0){\epsfig{file=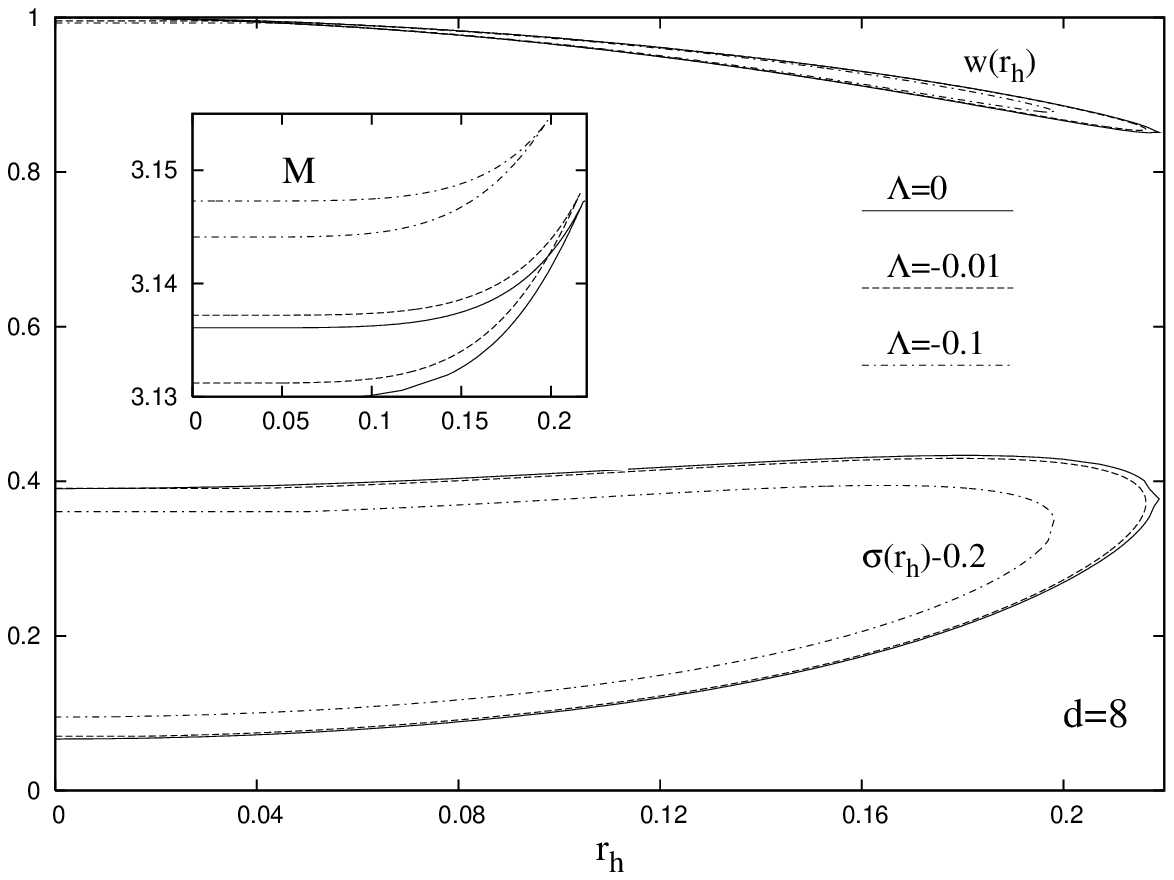,width=16cm}}
%\caption{a}
\end{picture}
\\
\\
\\
{\small {\bf Figure 13.}} 
The value of the gauge field function at the horizon $w(r_h)$ and
$\sigma(r_h)$, the value of the metric function $\sigma$ at the horizon, 
as well as the mass parameter $M$  
are shown as functions of the event horizon radius $r_h$ for
the $p=1,~2$   black hole solutions in eight dimensions with 
$\alpha^2=0.002$ and several values of $\Lambda$. 

%%%%%%%%%%%%%%%%%%%%%%%%%
\newpage
\setlength{\unitlength}{1cm}

\begin{picture}(16,16)
\centering
\put(-1,0){\epsfig{file=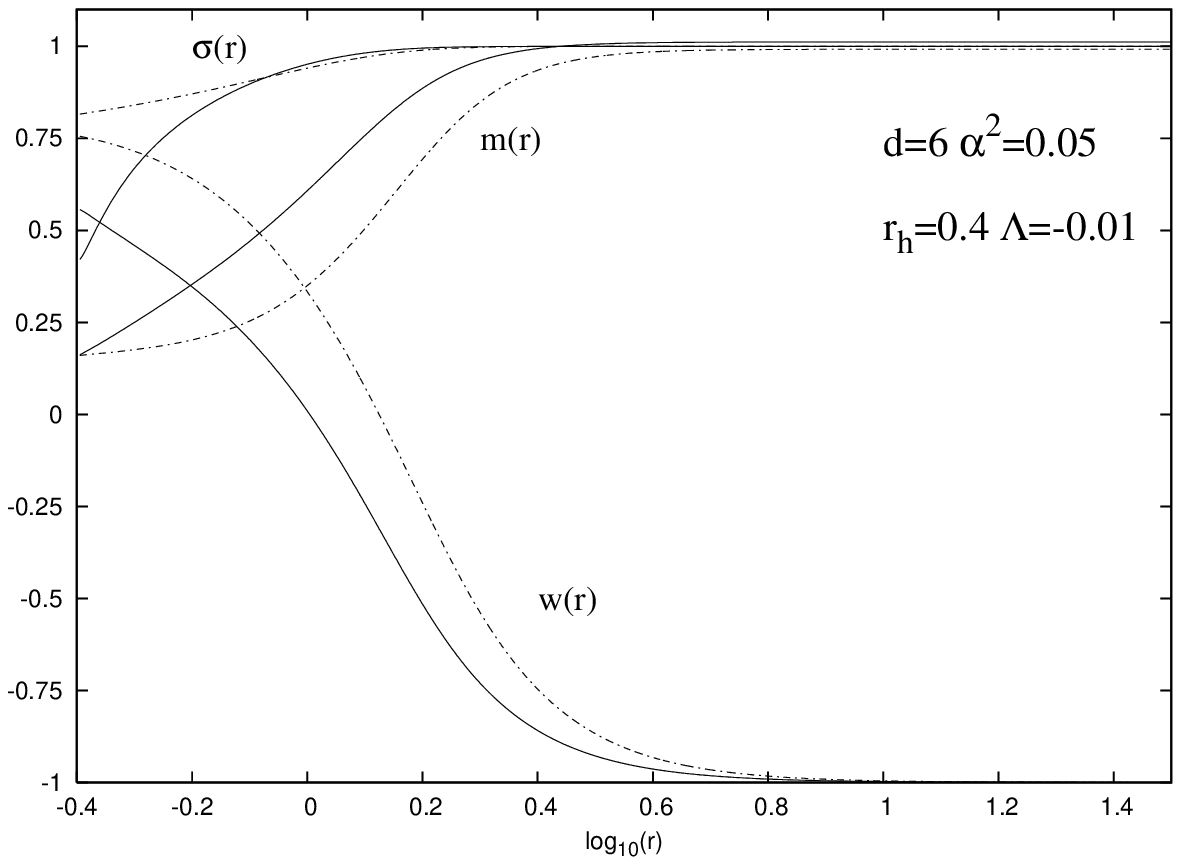,width=16cm}}
%\caption{a}
\end{picture}
\\
\\
\\
{\small {\bf Figure 14.}} 
The profiles of the functions $m(r),~\sigma(r)$ and $w(r)$  
are plotted as functions of the radius
for typical $d=6$ black hole solutions in a  
EYM theory with $p=1,2$ terms and $\alpha^2=0.05,~\Lambda=-0.01$.

%%%%%%%%%%%%%%%%%%%%%%%%
\newpage
\setlength{\unitlength}{1cm}

\begin{picture}(16,16)
\centering
\put(-1,0){\epsfig{file=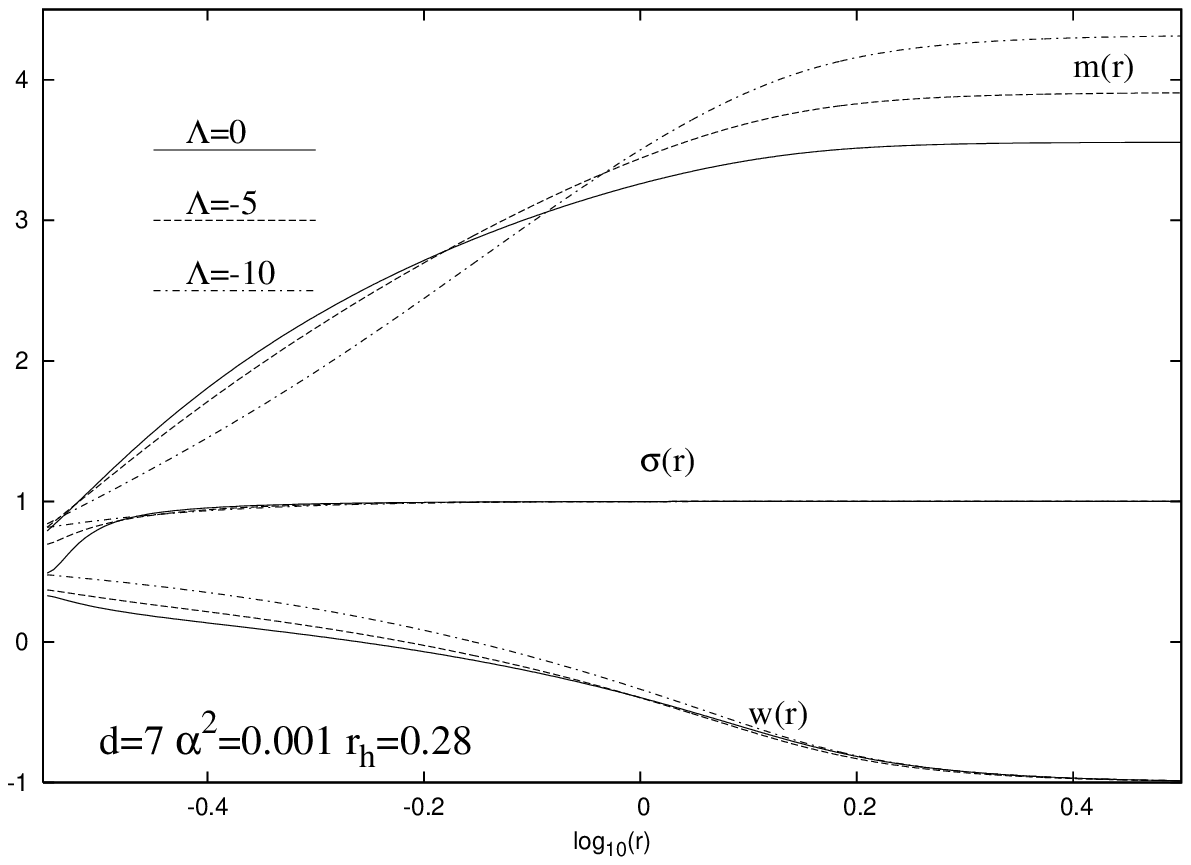,width=16cm}}
%\caption{a}
\end{picture}
\\
\\
\\
{\small {\bf Figure 15.}} 
Typical black hole solutions of  $p=1,2$ EYM theory in seven dimensions
with $\alpha^2=0.001,~r_h=0.28$
are plotted as functions of the radius for several values of $\Lambda$.

\end{document}